\documentclass[a4paper,twocolumn,showpacs,superscriptaddress,amsmath,amssymb,aps,prl,preprintnumbers]{revtex4-2}
\usepackage{bbm}
\usepackage{dsfont}
\usepackage{amsmath}
\usepackage{verbatim}
\usepackage{graphicx}
\usepackage{float}
\usepackage{braket,amsmath}
\usepackage[dvipsnames, svgnames, x11names]{xcolor}
\usepackage{array}
\usepackage{bibunits}
\usepackage{braket}
\usepackage{mathrsfs}
\usepackage{amsmath}
\usepackage{graphicx}
\usepackage{dcolumn}
\usepackage{bm}
\usepackage{hyperref}
\usepackage[capitalise]{cleveref}
\usepackage{upgreek}
\crefname{section}{Sec.}{Secs.}
\crefname{appendix}{App.}{Apps.}

\begin{document}

\emph{}
\title{Cooperative Squeezing of Internal and Collective Spins in an Atomic Ensemble}

\author{Youwei Zhang}%
\affiliation{%
Department of Physics, State Key Laboratory of Surface Physics and Key
Laboratory of Micro and Nano Photonic Structures (Ministry of Education),
Fudan University, Shanghai 200433, China
}%
\author{Shenchao Jin}%
\affiliation{%
CAS Key Laboratory of Quantum Optics and Aerospace Laser Technology and Systems Department, Shanghai Institute of Optics and Fine Mechanics, Chinese Academy of Sciences, Shanghai 201800, China
}
\affiliation{%
University of Chinese Academy of Sciences, Beijing 100049, China
}
\author{Junlei Duan}%
\affiliation{%
Department of Physics, State Key Laboratory of Surface Physics and Key
Laboratory of Micro and Nano Photonic Structures (Ministry of Education),
Fudan University, Shanghai 200433, China
}

\author{Klaus Mølmer}%
\affiliation{%
Niels Bohr Institute, University of Copenhagen, Jagtvej 155 A, DK 2200 Copenhagen, Denmark
}%

\author{Guiying Zhang}%
\email{guiyinzhang3619@zjut.edu.cn}
 \affiliation{%
School of Physics, Zhejiang University of Technology, Hangzhou 310023, China
}%
\author{ Mingfeng Wang}
\email{mfwang@wzu.edu.cn}
\affiliation{%
 Department of Physics, Wenzhou University, Zhejiang 325035, China
}%
\author{Yanhong Xiao}
\email{yxiao@fudan.edu.cn}
\affiliation{%
Department of Physics, State Key Laboratory of Surface Physics and Key
Laboratory of Micro and Nano Photonic Structures (Ministry of Education),
Fudan University, Shanghai 200433, China
}%
\affiliation{%
State Key Laboratory of Quantum Optics Technologies and Devices,
Institute of Laser Spectroscopy, Shanxi University, Taiyuan, Shanxi 030006, China
}%
\affiliation{%
Collaborative Innovation Center of Extreme Optics, Shanxi University, Taiyuan, Shanxi 030006, China
}%


\begin{abstract}
Creating highly spin-squeezed states for quantum metrology surpassing the standard quantum limit is a topic of great interest. Spin squeezing has been achieved by either entangling different atoms in an ensemble, or by controlling the multilevel internal spin state of an atom. Here, we experimentally demonstrate combined internal and collective spin squeezing in a hot atomic ensemble with $\sim 10^{11}$ rubidium atoms. By synergistically combining these two types of squeezing and carefully aligning their squeezing quadratures, we have achieved a metrologically relevant spin squeezing of $-6.21\pm0.84$ dB, significantly outperforming the results obtained by utilizing either type of squeezing alone in our system. Our approach provides a new perspective on fully harnessing the degrees of freedom inherent in quantum states of an atomic ensemble.
\end{abstract}
\maketitle

Squeezed spin states (SSSs) of atomic ensembles can have smaller quantum fluctuations in certain directions than coherent spin states \cite{PhysRevA.47.5138,PhysRevA.46.R6797,physics.reports1} and are currently attracting particular attention in various contexts. They are highly multipartite entangled states \cite{RevModPhys.81.865,RevModPhys.90.035005} that enable parameter sensing with precision beyond the standard quantum limit (SQL) set by the Heisenberg uncertainty relation \cite{RevModPhys.90.035005} and thus have direct applications in quantum metrology and quantum information sciences, such as atomic clocks \cite{pedrozo2020entanglement, robinson2024direct}, atom interferometers \cite{greve2022entanglement, malia2022distributed}, atomic magnetometers \cite{PhysRevLett.104.093602,PhysRevLett.113.103004,PhysRevLett.109.253605}, and continuous-variable quantum information processing \cite{RevModPhys.77.513}.
\begin{figure*}[htbp]
	\centering
	\includegraphics[width=2\columnwidth]{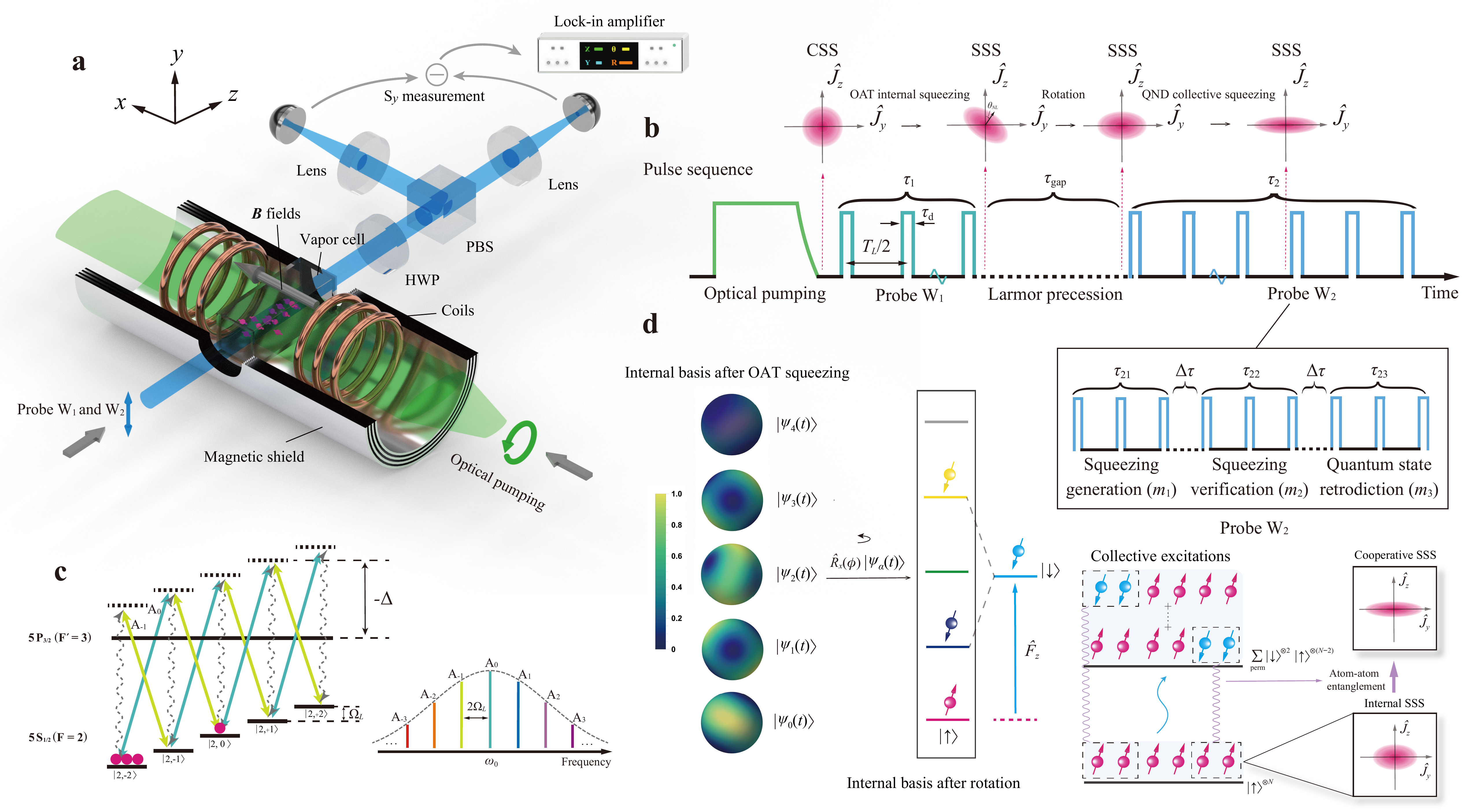}
	\caption{
(a) Experimental setup schematics. The $^{87} \mathrm{Rb}$ atoms are contained in a $7 \times 7 \times 20 \ \rm{mm}$ vapor cell placed inside a four-layer magnetic shield and coils producing a bias magnetic field along the $x$ direction. The $\sigma^{-}$-polarized pump and repump lasers both propagate along the $x$ direction, with the former tuned to the $D1$ transition $5S_{1/2}, F=2 \rightarrow 5P_{1/2}, F'=2$, and the latter to the $D2$ transition $5S_{1/2}, F=1 \rightarrow 5P_{3/2}, F'=2$. The two probe lasers propagate along the $z$ direction, with linear $y$ polarization. The probe W$_{1}$ is blue detuned by $1.6$ GHz from the $D2$ transition $5S_{1/2}, F=2 \rightarrow 5P_{3/2}, F'=3$, and the probe W$_{2}$ is blue detuned by $2.5$ GHz from $5S_{1/2}, F=2 \rightarrow 5P_{3/2}, F'=3$. The Stokes component $\hat S_{y}$ of the probe W$_{2}$ is detected via balanced homodyne detection. HWP, half-wave plate; PBS, polarization beam splitter. (b) Pulse sequence. Atoms are first prepared in the CSS by optical pumping, then interact with two stroboscopic pulses. The first probe W$_{1}$ creates the internal squeezing and the second probe W$_{2}$ produces collective squeezing. The probe W$_{2}$ consists of three parts: the first part creates the spin squeezing, the second part verifies the spin squeezing, and the third part further retrodicts the spin state. The time interval $\Delta \tau = 0.31$ ms between the three probe periods is set to prevent signal correlation due to the lock-in amplifier. (c) Atomic level scheme of the atom-light interactions. The stroboscopic pulse in the frequency domain can be viewed as a frequency comb (see Supplemental Material \cite{sm}) (right). The central carrier and the first sideband drive the resonant two-photon Raman transitions between magnetic sublevels. Dashed arrows represent the scattering of Stokes and anti-Stokes photons, which have the same frequency. (d) Pictorial representation of the cooperative spin squeezing. Left: Bloch sphere representation of the internal state due to OAT evolution. Middle: the five time-dependent internal states of a single atom after an $x$-axis rotation by angle  $\phi$ and the effective spin transition $\ket{\uparrow} \rightarrow\ket{\downarrow}$ (see text). Right: the QND detection process creates pairwise entanglement between atoms, which enhances the overall spin squeezing.
}
\label{fig1}
\end{figure*}

To date, studies on spin squeezing have mainly focused on establishing entanglement between different atoms within the ensemble by means of ``collective squeezing" through approaches such as Hamiltonian evolution \cite{PhysRevLett.104.073602,Nature60,Nature61} or quantum nondemolition (QND) measurements \cite{PhysRevLett.85.1594, appel2009mesoscopic,PhysRevLett.104.073604}. Each individual atom in the ensemble is normally treated as a qubit, with only two internal atomic levels participating in squeezed-state engineering \cite{NP23}. However, an atom often has more than two sublevels, proposed as ``qudits" \cite{PhysRevLett.99.163002} for quantum computing. By control of the qudit, squeezing of individual spins, i.e., ``internal squeezing" was realized in atomic ensembles~\cite{PhysRevLett.99.163002,PhysRevLett.101.073601}. A natural question then arises: could the combination of collective squeezing and internal squeezing enhance the overall spin-squeezing level? Theoretical studies predict that appropriate control of the internal and collective states could lead to the enhancement of overall squeezing \cite{PhysRevA.81.032314,PhysRevLett.109.173603}, but experimental studies are still lacking.

Here, we  present experimental demonstration of the cooperative integration of collective and internal spin squeezing in an atomic ensemble, resulting in an overall higher squeezing level than that of any of the two single types of squeezing. By judiciously engineering the system Hamiltonian and accurately controlling the spin state, we can combine the atom-atom entanglement and internal squeezing in one system in a manner that increases the achievable metrological gain. The obtained total squeezing is $-6.21\pm0.84$ dB in an ensemble of $N\sim 10^{11}$ rubidium atoms. Our method paves the way to use the qudit-subsystem control to enhance the total useful spin squeezing of a quantum system, which is applicable to various quantum platforms such as cold atoms \cite{PhysRevA.79.043815, appel2009mesoscopic, PhysRevA.104.023710}, Rydberg atoms \cite{PhysRevA.86.023845, PhysRevLett.112.103601}, and trapped ions \cite{gilmore2021quantum}.

As shown in \cref{fig1}(a), our quantum system involves a large $^{87} \mathrm{Rb}$ atom ensemble contained in a glass cell. The atomic states of concern are the five Zeeman substates with $m_F\in\{-2,-1,0,1,2\}$  of the $F=2$ manifold of the $5^2S_{1/2}$ ground state. These states comprise the relevant internal qudit subsystem. The atoms are initialized in the $\ket{\psi_0}$ ($\ket{\psi_\alpha}\equiv\ket{F=2,m_F=\alpha-2}$) sublevel, forming a coherent spin state (CSS) polarized along the quantization axis $\hat{\mathbf{x}}$. Then two $y$-polarized light pulses W$_{1}$ and W$_{2}$ with different center frequencies are sent into the ensemble sequentially to create internal and collective squeezing, respectively [\cref{fig1}(a)]. In the weak excitation limit, the atomic excited states can be adiabatically eliminated to yield \cite{sm}
\begin{eqnarray}
\hat H =  - \frac{1}{3}{\chi _2}\Phi \sum\nolimits_{i = 1}^N {\hat F_y^{(i)2}}  + {\chi _2}{\hat S_y}{\hat J_y} + {\chi _1}{\hat S_z}{\hat J_z},\label{eq1_1}
\end{eqnarray}
where $\chi_{1,2}$ denote the coupling constants and $\Phi$ represents the photon flux. $\hat S_{x,y,z}$ are the Stokes operators for light, $\hat F_{x,y,z}^{(i)}$ are the single-spin angular momentum operators of the $i$th atom, and ${\hat J_{x,y,z}} = \sum\nolimits_{i = 1}^N {\hat F_{x,y,z}^{(i)}}$ are the collective spin angular momentum operators. The first nonlinear term ${\hat H_{0}}= -\frac{1}{3}\chi_{2}{\Phi}{\hat{F}^{2}_{y}}$, known as the one-axis twisting (OAT) interaction \cite{PhysRevA.47.5138}, acts coherently on each atom, responsible for the internal squeezing; the third term is the well-known QND interaction, which will be employed to create collective squeezing.

When the probe light (W$_1$) is near resonance with the atomic transition, the OAT term in Eq.(\ref{eq1_1}) contributes significantly, while the other two terms have only negligible effects (see Supplemental Material \cite{sm}).
To conveniently clarify the core physics, let us for the moment neglect the atom-light interaction terms. Then, starting from the CSS $\ket{\psi_0}$, each atom evolves to $\ket{\psi _0(t)}$ ($\ket{\psi _{\alpha}(t)} \equiv {e^{ - it\hat H_0}} {\ket{ \psi_\alpha}}$), which is an internal SSS with a squeezing direction varying over time \cite{PhysRevA.47.5138}. This state will be rotated by an angle $\phi$ due to Larmor precession in $\bm{\mathit{B}}$, yielding $\ket{\uparrow}\equiv \hat{R}_x(\phi)\ket{\psi_0(t)}=\sum_{k=0}^{2}c_{2k}\ket{\psi_{2k}}$ with ${\hat{R}_x(\phi)= e^{ - i\phi \hat F_x}}$ and $c_{\alpha}=\braket{\psi_{\alpha}|\uparrow}$. To create collective squeezing, a ``far-off resonant" probe light (W$_2$) is sent through the ensemble, leading to $\chi_2\ll\chi_1$ and thus $\hat H\simeq\chi_1 \hat S_z\hat J_z$, which entangles light with atoms and, therefore, measuring $\hat S_y$ of the outgoing light squeezes $\hat J_z$ \cite{physics.reports1}. The coherent QND dynamics couples the internal SSS to an orthogonal state ${{\hat F}_z}\ket{\uparrow}=\Delta  F_z\ket{\downarrow}$, where $\ket{\downarrow}  \equiv \hat{R}_x(\phi)({c_{10}}\ket{ {{\psi _1}(t)} }  + {c_{30}}\ket{ {{\psi _3}(t)} } )/\Delta {{\hat F}_z}=\sum_{k=0}^{1}d_{2k+1}\ket{\psi_{2k+1}}$ with ${c}_{\alpha 0}=\bra{{\psi}_{\alpha}( t )} \hat{R}_x(-\phi)\hat{F}_{z} \hat{R}_x(\phi)\ket{{\psi}_{0}( t )}$, $d_{\alpha}=\braket{\psi_{\alpha}|\downarrow}$ and the normalization constant $\Delta  F_z$ being the standard deviation of $\hat F_z$ in the state $\ket{\uparrow}$ \cite{PhysRevA.81.032314}. Following the approach in \cite{PhysRevLett.109.173603}, for weak coupling limit ($\chi_1\ll 1$) and measurement outcome $S_y=0$, the ensemble state collapses to
\begin{eqnarray}
\ket{ {{\Psi _A}(t)} }  \approx {\ket{\uparrow} ^{ \otimes N}} - \frac{\tilde{\kappa}^2(\Delta F_z)^2}{4N}\sum\limits_{\rm{perm}} \ket{\downarrow}^{ \otimes 2}{\ket{\uparrow}^{ \otimes (N - 2)}},
\label{eq2_1}
\end{eqnarray}
where $\Tilde{\kappa}^{2} = N \Phi \chi^{2}_1 T$ with $T$ being the pulse duration. Spin-pair excitations from $\ket{\uparrow}$ to $\ket{\downarrow}$ are created, resulting in the pairwise entanglement that gives rise to collective squeezing. From Eq. (\ref{eq2_1}), one may derive the variance $(\Delta  J_z)^2\approx N(\Delta F_z)^2 -\tilde{\kappa}^2N(\Delta F_z)^4$, indicating that by appropriately adjusting $\phi$ to align the internal-state squeezing direction along the $z$ axis (minimizing $\Delta  F_z$), both internal and collective squeezing contribute to the total spin squeezing.

The full Hamiltonian (\ref{eq1_1}) induces both internal and collective evolution. However, the collective evolutions induced by the atom-light interaction terms only degrade the ensemble squeezing created by $\hat H_{0}$ and slightly alter its squeezing direction \cite{sm}. Therefore, the ensemble squeezing produced by W$_1$, quantified by the squeezing coefficient $\xi_{NL}^{2}$, is mainly contributed by the internal OAT evolution.
A measurement of W$_2$'s $\hat S_y$ component
and subsequent feedback of the measurement outcomes
result in the overall squeezing coefficient \cite{sm}
\begin{eqnarray}
\xi_{{\rm{tot}}}^2 = {\xi_{NL}^2}/\left({ {1 + {{\tilde \kappa}^2 \xi_{NL}^2}} }\right).\label{eq2}
\end{eqnarray}
This is an improvement over the conventional QND squeezing coefficient $\xi_{\rm{QND}}^2 = 1/(1 + {\tilde \kappa^2})$, imposed solely by the internal squeezing ($\xi_{NL}^2<1$).  The resulting squeezing coefficient of the combined scheme, however, is larger (worse) than the product of the internal and collective squeezing coefficient. This is because the
internal squeezing reduces the QND coupling strength by
a factor $\xi_{NL}^2$, which decreases the efficiency of the QND
measurement \cite{PhysRevA.81.032314}. In addition, the competition between coherent interaction and spontaneous emission limits squeezing to $\xi_{{\rm{tot}}}^2 =2/\sqrt{1/\xi_{{NL}}^2+\alpha_0}$, with $\alpha_0$ being the optical depth \cite{sm}. Apparently, the internal squeezing alters the populations and coherences of the Zeeman sublevels which effectively enhances the optical depth and thereby enables surpassing the squeezing limit of the QND scheme. The observed spin squeezing is limited by photon scattering and the associated Raman processes, which can be explained by a detailed noise budget (see Supplemental Material \cite{sm}) taking into account this effect.

\begin{figure}[tp]
  \centering
  \includegraphics[scale=0.29]{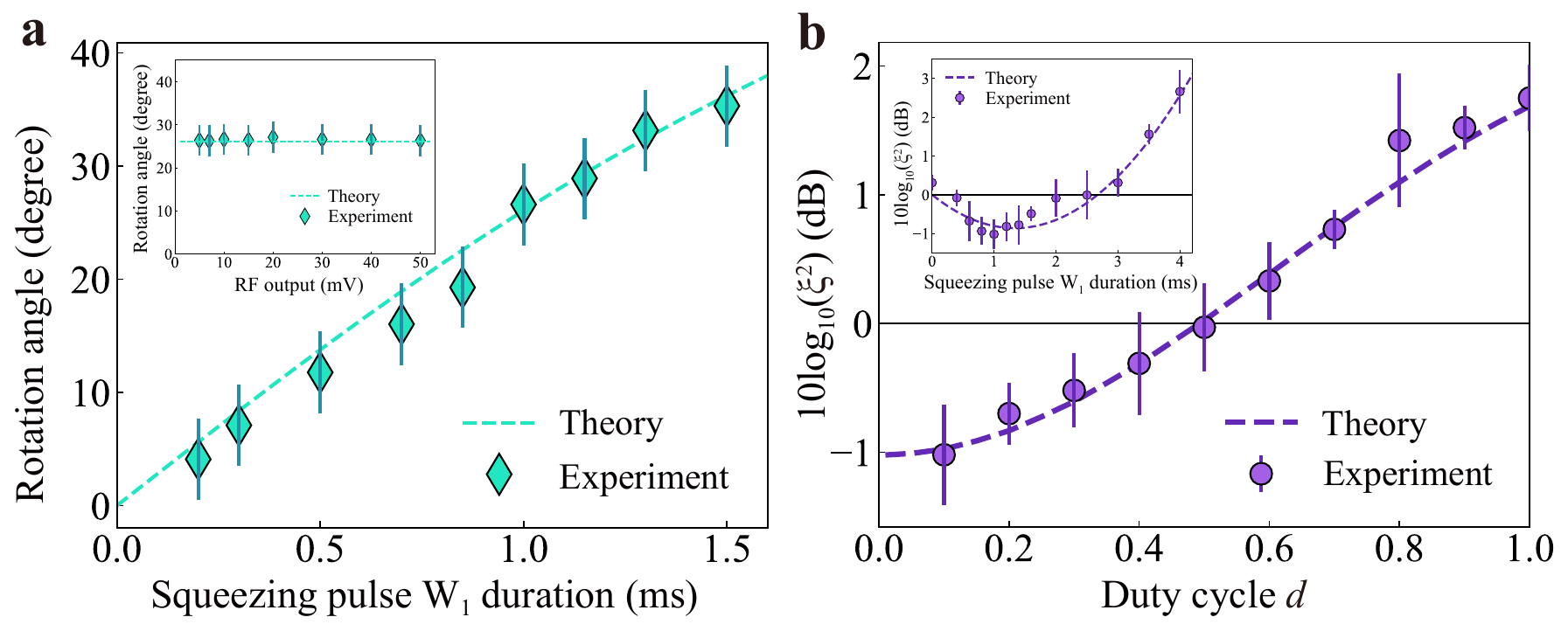}
  \caption{(a) Rotation angle of the transverse mean spin versus the squeezing pulse duration of W$_{1}$ for $d=0.1$ and mean power $0.8$ mW. Inset: rotation angle versus the amplitude of the rf field (proportional to the created mean spin value) with the W$_{1}$ pulse duration of $1.0$ ms. With a standard deviation of about $\pm0.3^\circ$ (not shown), each data point is the average of five identical experiments each consisting of 1000 repeated measurements. The accuracy of the rotation angle is technically constrained by the minimum pulse delay time $0.02$ $\upmu$s, which is equal to $1\%$ of a Larmor period, causing an error of $\pm3.6^\circ$ (shown). (b) Internal spin squeezing versus duty cycle $d$ with pulse duration $\tau_{1}=1.0$ ms and mean laser power $0.8$ mW. Inset: internal spin squeezing versus pulse duration of W$_{1}$ with $d=0.1$, showing that there exists an optimal squeezing. The data in the main figure are pulse-duration optimized. The error bar for each data point represents the standard deviation of five identical experiments each consisting of 10000 pulse (optical pumping $+$ probe) cycles. The black solid line denotes the SQL.
  }
  \label{fig2}
\end{figure}

 In our experiment, the paraffin-coated atomic vapor cell is placed inside a magnetic shield to screen out ambient magnetic fields. A bias magnetic field $\bm{\mathit{B}}$ is applied along the $x$ direction to hold the large collective spin, which also causes Larmor precession at a frequency of $\Omega_{L} \approx 2\pi \times 500$ kHz. Atoms are first optically pumped to $\ket{\psi_0}$, with a measured degree of polarization of
about $97.4\%$, which gives about $7\%$ excess noise in spin variance compared to a perfect CSS. Then two probe pulses, W$_{1}$ and W$_{2}$, are applied, both linearly $y$ polarized and propagating along the $z$ direction, but with different central frequencies. The probe W$_{1}$ drives a near-resonant transition leading to nonlinear evolution of the internal spin, while the probe W$_{2}$ induces a far-off-resonant Faraday-type QND interaction for spin readout and collective spin squeezing. Both W$_{1}$ and W$_{2}$ are stroboscopic pulses [see \cref{fig1}(b)] produced by acousto-optic modulators to evade quantum backaction \cite{PhysRevLett.133.173604} for implementation of the Hamiltonian (\ref{eq1_1}).

We first examine the effect of the nonlinear interaction ${\hat H_{0}}$. The squeezing direction of the internal OAT evolution varies with time, i.e., rotating around the $x$ axis \cite{PhysRevLett.105.193602}. This feature allows one to identify the OAT dynamics via monitoring the mean-value evolution of a displaced atomic state. After preparing the CSS we apply an rf magnetic field pulse with a frequency equal to the Larmor frequency (produced by a pair of transverse coils inside the magnetic shield) along the $z$ axis to create a displaced CSS with mean value $\langle\hat J_y\rangle= J_0,\langle\hat J_z\rangle=0$. Then, the probe W$_{1}$ is turned on to induce the $\hat H$ interaction, followed by a W$_{2}$ pulse to detect the mean values of $\hat J_{y,z}$ components. The $\hat H_{0}$ interaction causes a spin rotation, since $\langle\hat J_z(t)\rangle= \text{Re}J_{10}^z(t)J_0$, with $J_{10}^z(t)=\bra{{\psi}_{1}( t )} \hat{F}_{z} \ket{{\psi}_{0}( t )}$ \cite{sm}, which gives a rotation angle of $\arctan\langle\hat J_z\rangle/\langle\hat J_y\rangle$.  \cref{fig2}(a) plots the measurement results, indicating that the rotation angle increases with the time duration of W$_{1}$, in good agreement with the theoretical predictions. In addition, we observe that for a fixed coupling strength, different mean values of $\hat J_y$ yield almost the same rotation angle, which further confirms  that the qudit subsystem evolves according to ${\hat H_{0}}$ since $\langle\hat J_z\rangle/\langle\hat J_y\rangle=\text{Re}J_{10}^z$ is independent of $J_0$.

\begin{figure}[tp]
  \centering
  \includegraphics[scale=0.28]{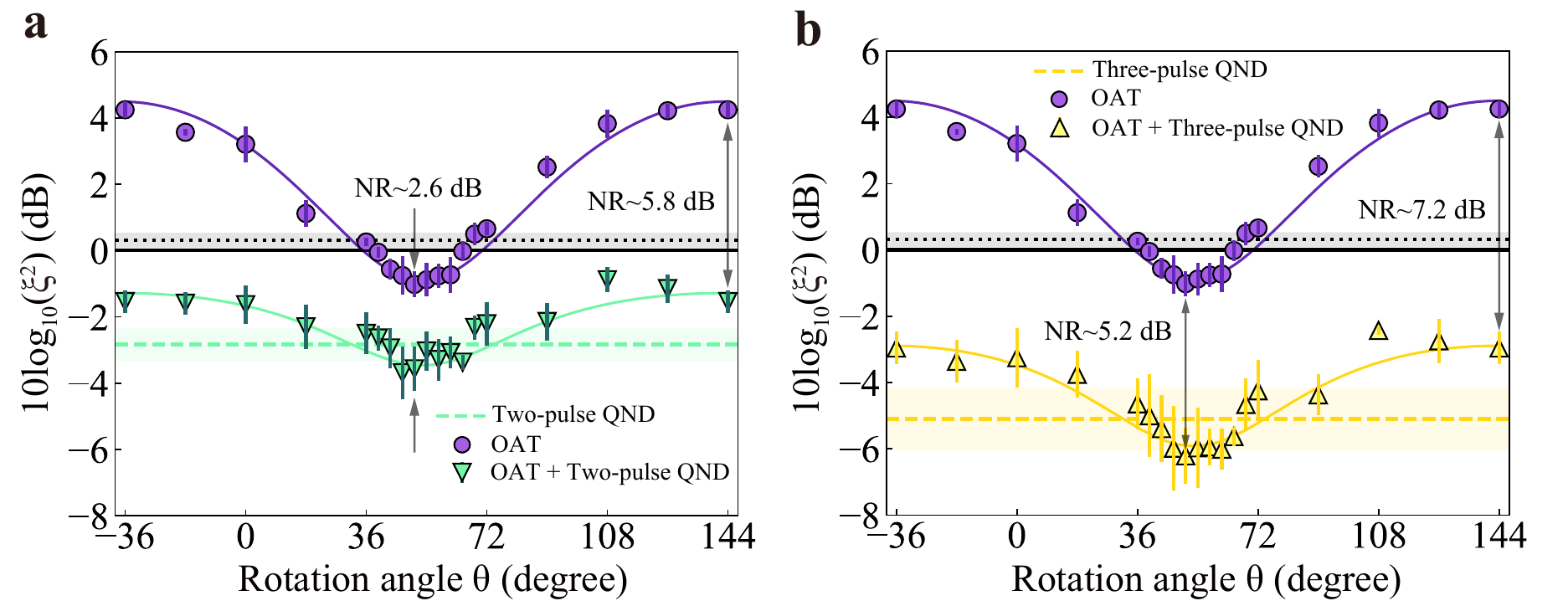}
  \caption{Spin squeezing versus rotation angle $\theta$ for $d=0.1$. The black dotted and solid lines in both (a) and (b) represent the noise level of the prepared CSS  ($0.31\pm0.20$ dB for 97.4$\%$ spin polarization), and the SQL (0 dB), respectively. The green dashed line ($-2.83\pm0.47$ dB) in (a) and the yellow dashed line ($-5.10\pm0.93$ dB) in (b) denote the maximal squeezing achieved solely by the two- and three-pulse QND schemes (without internal squeezing), respectively. The purple, green and yellow solid curves are the theoretical predictions. The mean power is $0.8$ and $1.0$ mW for the probe W$_{1}$ and the probe W$_{2}$, respectively. The error bar for each data point represents the standard deviation of five identical experiments, each consisting of 10000 cycles. The rotation angle is calibrated through the mean spin measurement, and its accuracy is also limited by the minimum pulse delay time. NR denotes the noise reduction caused by the QND measurement.
  }
  \label{fig3}
\end{figure}

Next, we confirm the generation of internal spin squeezing. As shown in \cref{fig1}(c), the internal OAT process established by using the stroboscopic probe W$_{1}$ drives the Raman transitions resonantly between magnetic sublevels with $\Delta m_F=2$, causing internal squeezing \cite{PhysRevLett.101.073601}. This can be characterized by the QND probe pulse W$_{2}$, which is also stroboscopic to evade backaction \cite{Nature.581.7807}. To quantify the squeezing we use the Wineland criterion \cite{PhysRevA.50.67} $\xi_{NL}^2 = e^{2T/T_1}  \mathrm{Var}(\hat{J}_{z})_\mathrm{SSS} /  \mathrm{Var}(\hat{J}_{z})_\mathrm{PNL}$, where $\mathrm{Var}(\hat{J}_z) = \langle \hat{J}^2_z \rangle - \langle \hat{J}_z \rangle^2$ and the prefactor $e^{2T/T_1}$ accounts for the decay of the macroscopic spin with the relaxation time $T_1 = 37~\mathrm{ms}$. $\mathrm{Var}(\hat{J}_{z})_\mathrm{PNL}$ represents the spin projection noise limit (PNL) \cite{NP23, Nature.581.7807}. \cref{fig2}(b) shows the squeezing versus the duty cycle $d$, indicating that smaller $d$ (more efficient OAT) gives larger squeezing. Since the transverse spin ellipse precesses around the $x$ axis due to the bias magnetic field, one can rotate any spin component in the $y$-$z$ plane to the $z$ direction simply by varying the time interval between the squeezing and readout pulses [that is, $\tau_{\rm{gap}}$ in \cref{fig1}(b)], enabling the observation of internal spin squeezing along different directions.
\cref{fig3}(a) plots the measured internal squeezing along different directions for $d=0.1$, showing that the maximal internal squeezing is $10\mathrm{log}_{10}(\xi_{NL}^2)=-1.02\pm0.39~\mathrm{dB}$, which includes the effect of the mean spin's shortening (about $0.23$ dB) contributed from both optical pumping and the coherent oversqueezing dynamics during OAT.

We now proceed to the combined squeezing of internal and collective spins. By adjusting the Larmor precession time $\tau_{\rm{gap}}$ following the W$_{1}$ pulse, one can direct the maximally squeezed spin component along the $z$ direction. Then, a stroboscopic probe W$_{2}$ is sent through the atomic sample to experience the QND interaction \cite{NP23, PhysRevLett.106.143601}, described by the input-output relation $\hat{S}_{y}^{\text{out}} = \hat{S}_{y}^{\text{in}} - \frac{1}{2} \chi \Phi  \hat{J}_{\text{opt}}^{\theta}$, where the superscript in (out) denotes the operators before (after) the interaction. Therefore the information of $\hat J_{\rm{opt}}^\theta$ can be obtained by measuring $\hat{S}_{y}^{\text{out}}$ of the W$_{2}$ pulse, further reducing the uncertainty of the squeezed spin component. The overall squeezing of the combined scheme can be obtained from the W$_{2}$ two-pulse measurement results $m_{1}$ and $m_{2}$ \cite{Nature.581.7807} [see \cref{fig1}(b)]. The variance of $m_{2}$ conditioned on $m_{1}$ is ${\text{Var}(m_{2}|m_{1})=V_2-C_{12}^{2}/V_1}$, where $V_i=\text{Var}(m_{i})$ and $C_{ij}=\text{Cov}(m_{i}, m_{j})$. 

The total degradation in the metrological gain due to spin shortening in the OAT and QND is about $0.54$ dB, which has been accounted for in the $\xi_{\rm{tot}}^2$ (see Supplemental Material \cite{sm}). Taking into account the macroscopic-spin decay, we achieve a combined total squeezing of $10\mathrm{log}_{10}(\xi_{\rm{tot}}^2)=-3.57\pm0.67$ dB, which is better than the squeezing $10\mathrm{log}_{10}(\xi_{\rm{QND}}^2)=-2.83\pm0.47$ dB attained by the QND scheme alone [\cref{fig3}(a)]. The combined squeezing, however, is slightly smaller than the direct decibel sum of the internal and collective squeezing obtained when they were performed independently, which is consistent with the theoretical prediction given by Eq.(\ref{eq2}). We note that the excess noise caused by the imperfect polarization of CSS is included in the analysis.

The overall squeezing can be further enhanced by retrodiction through later  measurements based on the past quantum state (PQS) formalism \cite{PhysRevLett.102.250403, PhysRevLett.111.160401,PhysRevA.96.062131}. In contrast to the conventional prediction-based two-pulse QND, the PQS uses a three-pulse QND and can make better estimates for the unknown outcome of any measurement at $t$, conditioned on the information obtained both before and after $t$ \cite{PhysRevLett.111.160401, Nature.581.7807}.
To perform the PQS, after the measurement $\tau_{22}$ we continue to detect the probe W$_{2}$ for another duration of $\tau_{23}$ [see \cref{fig1}(b)]. \cref{fig3}(b) shows the squeezing produced by the three-pulse QND scheme along different directions. A maximal total squeezing of $-6.21\pm0.84$ dB is obtained, which is, to the best of our knowledge, the highest squeezing achieved so far in a hot atomic ensemble. We emphasize an interesting fact that the three-pulse QND integrates better with the internal squeezing than the two-pulse QND; i.e., the total squeezing in decibel is closer to that of the sum of the separate internal and collective squeezing, because the third QND pulse is further away in time from W$_1$ and sees a less-reduced effective coupling strength by the internal squeezing. The retrodictive squeezing can be applied to detect an unknown temporary phase shift during $\tau_{22}$,  as demonstrated in our previous work on predictive and retrodictive QND spin squeezing \cite{Nature.581.7807}, and future continuous operation of OAT and QND will enable tracking a continuous time-varying rf magnetic field as discussed in \cite{PhysRevA.81.032314, PhysRevA.108.043722}.

The maximal achievable enhancement of squeezing by OAT is mainly limited by the dimension of the Hilbert space of the qudit subsystem. Larger internal squeezing is expected by using larger-$F$ spins, such as cesium ($F=4$) \cite{PhysRevLett.101.073601} or dysprosium ($F=8$) \cite{PhysRevLett.122.173601} atoms. Another possibility for improvements is preparing the atoms in a state with large quantum fluctuations \cite{PhysRevLett.109.173603}, which strengthens the Faraday interaction and thus may increase the collective squeezing. As is evident from \cref{fig3}, large spin fluctuations (along the antisqueezing spin direction) enhance our ability to achieve greater noise reduction, e.g., even reaching up to $7.2$ dB with the three-pulse scheme. However, overall less squeezing is observed due to higher initial spin uncertainty in the internal state and inefficiencies in the QND squeezing. Future work may explore the use of more efficient two-axis-countertwisting collective squeezing \cite{PhysRevA.96.013823,PhysRevA.47.5138,luo2024hamiltonianengineeringcollectivexyz, PhysRevLett.105.193602} where increasing the noise fluctuations of the internal state may have substantial effect on the maximal achievable squeezing \cite{PhysRevLett.109.173603}.

In conclusion, we have achieved cooperative squeezing of two different types of spins---internal spin for a single atom and collective spin of the atomic ensemble, respectively, by deterministic light-induced nonlinear Hamiltonian evolution and by QND measurement. Through coherent control of these two types of squeezing, we are able to integrate them constructively to enhance the overall squeezing up to about $-6.21$ dB, outperforming the results obtained by applying either type of spin squeezing alone in the system. Our analysis and scheme can be generalized to the overall optimal control of the qudit ensemble systems, by integrating arbitrary internal states \cite{PhysRevLett.99.163002} including non-Gaussian \cite{PhysRevLett.122.173601} and cat states \cite{yang2025minute} with atom-atom entanglement \cite{li2023improving,colombo2022time, Nature61,liu2022nonlinear} techniques, thus opening new avenues in quantum metrology \cite{RevModPhys.90.035005}, fundamental physics \cite{PhysRevLett.129.083001}, quantum simulation \cite{PhysRevLett.124.230501}, quantum memory \cite{PhysRevLett.101.073601}, and quantum information science \cite{RevModPhys.77.513}. For example, it is possible to transfer the squeezing to the clock transitions \cite{pedrozo2020entanglement} by suitable composite pulses \cite{PhysRevLett.99.163002} and to employ ``amplification-deamplification" protocols \cite{25ds-9724}.

\emph{Acknowledgments}---This work is supported by the Quantum Science and Technology-National Science and Technology Major Project (Grant No. 2023ZD0300900), STCSM24LZ1400400, the Natural Science Foundation of China (Grants No. 12027806 and No. 12161141018), Fund for Shanxi ``1331 Project," the Zhejiang Provincial Natural Science Foundation of China (Grant No. LMS25A040004), and the Danish National Research Foundation (Center of Excellence “Hy-Q,” Grant No. DNRF139).

\bibliography{main}

\providecommand{\noopsort}[1]{}\providecommand{\singleletter}[1]{#1}%
\begin{thebibliography}{55}%
\makeatletter
\providecommand \@ifxundefined [1]{%
 \@ifx{#1\undefined}
}%
\providecommand \@ifnum [1]{%
 \ifnum #1\expandafter \@firstoftwo
 \else \expandafter \@secondoftwo
 \fi
}%
\providecommand \@ifx [1]{%
 \ifx #1\expandafter \@firstoftwo
 \else \expandafter \@secondoftwo
 \fi
}%
\providecommand \natexlab [1]{#1}%
\providecommand \enquote  [1]{``#1''}%
\providecommand \bibnamefont  [1]{#1}%
\providecommand \bibfnamefont [1]{#1}%
\providecommand \citenamefont [1]{#1}%
\providecommand \href@noop [0]{\@secondoftwo}%
\providecommand \href [0]{\begingroup \@sanitize@url \@href}%
\providecommand \@href[1]{\@@startlink{#1}\@@href}%
\providecommand \@@href[1]{\endgroup#1\@@endlink}%
\providecommand \@sanitize@url [0]{\catcode `\\12\catcode `\$12\catcode `\&12\catcode `\#12\catcode `\^12\catcode `\_12\catcode `\%12\relax}%
\providecommand \@@startlink[1]{}%
\providecommand \@@endlink[0]{}%
\providecommand \url  [0]{\begingroup\@sanitize@url \@url }%
\providecommand \@url [1]{\endgroup\@href {#1}{\urlprefix }}%
\providecommand \urlprefix  [0]{URL }%
\providecommand \Eprint [0]{\href }%
\providecommand \doibase [0]{https://doi.org/}%
\providecommand \selectlanguage [0]{\@gobble}%
\providecommand \bibinfo  [0]{\@secondoftwo}%
\providecommand \bibfield  [0]{\@secondoftwo}%
\providecommand \translation [1]{[#1]}%
\providecommand \BibitemOpen [0]{}%
\providecommand \bibitemStop [0]{}%
\providecommand \bibitemNoStop [0]{.\EOS\space}%
\providecommand \EOS [0]{\spacefactor3000\relax}%
\providecommand \BibitemShut  [1]{\csname bibitem#1\endcsname}%
\let\auto@bib@innerbib\@empty
\bibitem [{\citenamefont {Kitagawa}\ and\ \citenamefont {Ueda}(1993)}]{PhysRevA.47.5138}%
  \BibitemOpen
  \bibfield  {author} {\bibinfo {author} {\bibfnamefont {M.}~\bibnamefont {Kitagawa}}\ and\ \bibinfo {author} {\bibfnamefont {M.}~\bibnamefont {Ueda}},\ }\bibfield  {title} {\bibinfo {title} {Squeezed spin states},\ }\href {https://doi.org/10.1103/PhysRevA.47.5138} {\bibfield  {journal} {\bibinfo  {journal} {Phys. Rev. A}\ }\textbf {\bibinfo {volume} {47}},\ \bibinfo {pages} {5138} (\bibinfo {year} {1993})}\BibitemShut {NoStop}%
\bibitem [{\citenamefont {Wineland}\ \emph {et~al.}(1992)\citenamefont {Wineland}, \citenamefont {Bollinger}, \citenamefont {Itano}, \citenamefont {Moore},\ and\ \citenamefont {Heinzen}}]{PhysRevA.46.R6797}%
  \BibitemOpen
  \bibfield  {author} {\bibinfo {author} {\bibfnamefont {D.~J.}\ \bibnamefont {Wineland}}, \bibinfo {author} {\bibfnamefont {J.~J.}\ \bibnamefont {Bollinger}}, \bibinfo {author} {\bibfnamefont {W.~M.}\ \bibnamefont {Itano}}, \bibinfo {author} {\bibfnamefont {F.~L.}\ \bibnamefont {Moore}},\ and\ \bibinfo {author} {\bibfnamefont {D.~J.}\ \bibnamefont {Heinzen}},\ }\bibfield  {title} {\bibinfo {title} {Spin squeezing and reduced quantum noise in spectroscopy},\ }\href {https://doi.org/10.1103/PhysRevA.46.R6797} {\bibfield  {journal} {\bibinfo  {journal} {Phys. Rev. A}\ }\textbf {\bibinfo {volume} {46}},\ \bibinfo {pages} {R6797} (\bibinfo {year} {1992})}\BibitemShut {NoStop}%
\bibitem [{\citenamefont {Ma}\ \emph {et~al.}(2011)\citenamefont {Ma}, \citenamefont {Wang}, \citenamefont {Sun},\ and\ \citenamefont {Nori}}]{physics.reports1}%
  \BibitemOpen
  \bibfield  {author} {\bibinfo {author} {\bibfnamefont {J.}~\bibnamefont {Ma}}, \bibinfo {author} {\bibfnamefont {X.~G.}\ \bibnamefont {Wang}}, \bibinfo {author} {\bibfnamefont {C.}~\bibnamefont {Sun}},\ and\ \bibinfo {author} {\bibfnamefont {F.}~\bibnamefont {Nori}},\ }\bibfield  {title} {\bibinfo {title} {Quantum spin squeezing},\ }\href {https://doi.org/https://doi.org/10.1016/j.physrep.2011.08.003} {\bibfield  {journal} {\bibinfo  {journal} {Phys. Rep.}\ }\textbf {\bibinfo {volume} {509}},\ \bibinfo {pages} {89} (\bibinfo {year} {2011})}\BibitemShut {NoStop}%
\bibitem [{\citenamefont {Horodecki}\ \emph {et~al.}(2009)\citenamefont {Horodecki}, \citenamefont {Horodecki}, \citenamefont {Horodecki},\ and\ \citenamefont {Horodecki}}]{RevModPhys.81.865}%
  \BibitemOpen
  \bibfield  {author} {\bibinfo {author} {\bibfnamefont {R.}~\bibnamefont {Horodecki}}, \bibinfo {author} {\bibfnamefont {P.}~\bibnamefont {Horodecki}}, \bibinfo {author} {\bibfnamefont {M.}~\bibnamefont {Horodecki}},\ and\ \bibinfo {author} {\bibfnamefont {K.}~\bibnamefont {Horodecki}},\ }\bibfield  {title} {\bibinfo {title} {Quantum entanglement},\ }\href {https://doi.org/10.1103/RevModPhys.81.865} {\bibfield  {journal} {\bibinfo  {journal} {Rev. Mod. Phys.}\ }\textbf {\bibinfo {volume} {81}},\ \bibinfo {pages} {865} (\bibinfo {year} {2009})}\BibitemShut {NoStop}%
\bibitem [{\citenamefont {Pezz\`e}\ \emph {et~al.}(2018)\citenamefont {Pezz\`e}, \citenamefont {Smerzi}, \citenamefont {Oberthaler}, \citenamefont {Schmied},\ and\ \citenamefont {Treutlein}}]{RevModPhys.90.035005}%
  \BibitemOpen
  \bibfield  {author} {\bibinfo {author} {\bibfnamefont {L.}~\bibnamefont {Pezz\`e}}, \bibinfo {author} {\bibfnamefont {A.}~\bibnamefont {Smerzi}}, \bibinfo {author} {\bibfnamefont {M.~K.}\ \bibnamefont {Oberthaler}}, \bibinfo {author} {\bibfnamefont {R.}~\bibnamefont {Schmied}},\ and\ \bibinfo {author} {\bibfnamefont {P.}~\bibnamefont {Treutlein}},\ }\bibfield  {title} {\bibinfo {title} {Quantum metrology with nonclassical states of atomic ensembles},\ }\href {https://doi.org/10.1103/RevModPhys.90.035005} {\bibfield  {journal} {\bibinfo  {journal} {Rev. Mod. Phys.}\ }\textbf {\bibinfo {volume} {90}},\ \bibinfo {pages} {035005} (\bibinfo {year} {2018})}\BibitemShut {NoStop}%
\bibitem [{\citenamefont {Pedrozo-Pe{\~n}afiel}\ \emph {et~al.}(2020)\citenamefont {Pedrozo-Pe{\~n}afiel}, \citenamefont {Colombo}, \citenamefont {Shu}, \citenamefont {Adiyatullin}, \citenamefont {Li}, \citenamefont {Mendez}, \citenamefont {Braverman}, \citenamefont {Kawasaki}, \citenamefont {Akamatsu}, \citenamefont {Xiao},\ and\ \citenamefont {Vuleti{\'c}}}]{pedrozo2020entanglement}%
  \BibitemOpen
  \bibfield  {author} {\bibinfo {author} {\bibfnamefont {E.}~\bibnamefont {Pedrozo-Pe{\~n}afiel}}, \bibinfo {author} {\bibfnamefont {S.}~\bibnamefont {Colombo}}, \bibinfo {author} {\bibfnamefont {C.}~\bibnamefont {Shu}}, \bibinfo {author} {\bibfnamefont {A.~F.}\ \bibnamefont {Adiyatullin}}, \bibinfo {author} {\bibfnamefont {Z.}~\bibnamefont {Li}}, \bibinfo {author} {\bibfnamefont {E.}~\bibnamefont {Mendez}}, \bibinfo {author} {\bibfnamefont {B.}~\bibnamefont {Braverman}}, \bibinfo {author} {\bibfnamefont {A.}~\bibnamefont {Kawasaki}}, \bibinfo {author} {\bibfnamefont {D.}~\bibnamefont {Akamatsu}}, \bibinfo {author} {\bibfnamefont {Y.}~\bibnamefont {Xiao}},\ and\ \bibinfo {author} {\bibfnamefont {V.}~\bibnamefont {Vuleti{\'c}}},\ }\bibfield  {title} {\bibinfo {title} {Entanglement on an optical atomic-clock transition},\ }\href {https://doi.org/10.1038/s41586-020-3006-1} {\bibfield  {journal} {\bibinfo  {journal} {Nature}\ }\textbf {\bibinfo {volume} {588}},\ \bibinfo {pages} {414} (\bibinfo {year}
  {2020})}\BibitemShut {NoStop}%
\bibitem [{\citenamefont {Robinson}\ \emph {et~al.}(2024)\citenamefont {Robinson}, \citenamefont {Miklos}, \citenamefont {Tso}, \citenamefont {Kennedy}, \citenamefont {Bothwell}, \citenamefont {Kedar}, \citenamefont {Thompson},\ and\ \citenamefont {Ye}}]{robinson2024direct}%
  \BibitemOpen
  \bibfield  {author} {\bibinfo {author} {\bibfnamefont {J.~M.}\ \bibnamefont {Robinson}}, \bibinfo {author} {\bibfnamefont {M.}~\bibnamefont {Miklos}}, \bibinfo {author} {\bibfnamefont {Y.~M.}\ \bibnamefont {Tso}}, \bibinfo {author} {\bibfnamefont {C.~J.}\ \bibnamefont {Kennedy}}, \bibinfo {author} {\bibfnamefont {T.}~\bibnamefont {Bothwell}}, \bibinfo {author} {\bibfnamefont {D.}~\bibnamefont {Kedar}}, \bibinfo {author} {\bibfnamefont {J.~K.}\ \bibnamefont {Thompson}},\ and\ \bibinfo {author} {\bibfnamefont {J.}~\bibnamefont {Ye}},\ }\bibfield  {title} {\bibinfo {title} {Direct comparison of two spin-squeezed optical clock ensembles at the $10^{-17}$ level},\ }\href {https://doi.org/10.1038/s41567-023-02310-1} {\bibfield  {journal} {\bibinfo  {journal} {Nat. Phys.}\ }\textbf {\bibinfo {volume} {20}},\ \bibinfo {pages} {208} (\bibinfo {year} {2024})}\BibitemShut {NoStop}%
\bibitem [{\citenamefont {Greve}\ \emph {et~al.}(2022)\citenamefont {Greve}, \citenamefont {Luo}, \citenamefont {Wu},\ and\ \citenamefont {Thompson}}]{greve2022entanglement}%
  \BibitemOpen
  \bibfield  {author} {\bibinfo {author} {\bibfnamefont {G.~P.}\ \bibnamefont {Greve}}, \bibinfo {author} {\bibfnamefont {C.}~\bibnamefont {Luo}}, \bibinfo {author} {\bibfnamefont {B.}~\bibnamefont {Wu}},\ and\ \bibinfo {author} {\bibfnamefont {J.~K.}\ \bibnamefont {Thompson}},\ }\bibfield  {title} {\bibinfo {title} {Entanglement-enhanced matter-wave interferometry in a high-finesse cavity},\ }\href {https://doi.org/10.1038/s41586-022-05197-9} {\bibfield  {journal} {\bibinfo  {journal} {Nature}\ }\textbf {\bibinfo {volume} {610}},\ \bibinfo {pages} {472} (\bibinfo {year} {2022})}\BibitemShut {NoStop}%
\bibitem [{\citenamefont {Malia}\ \emph {et~al.}(2022)\citenamefont {Malia}, \citenamefont {Wu}, \citenamefont {Mart{\'\i}nez-Rinc{\'o}n},\ and\ \citenamefont {Kasevich}}]{malia2022distributed}%
  \BibitemOpen
  \bibfield  {author} {\bibinfo {author} {\bibfnamefont {B.~K.}\ \bibnamefont {Malia}}, \bibinfo {author} {\bibfnamefont {Y.}~\bibnamefont {Wu}}, \bibinfo {author} {\bibfnamefont {J.}~\bibnamefont {Mart{\'\i}nez-Rinc{\'o}n}},\ and\ \bibinfo {author} {\bibfnamefont {M.~A.}\ \bibnamefont {Kasevich}},\ }\bibfield  {title} {\bibinfo {title} {Distributed quantum sensing with mode-entangled spin-squeezed atomic states},\ }\href {https://doi.org/10.1038/s41586-022-05363-z} {\bibfield  {journal} {\bibinfo  {journal} {Nature}\ }\textbf {\bibinfo {volume} {612}},\ \bibinfo {pages} {661} (\bibinfo {year} {2022})}\BibitemShut {NoStop}%
\bibitem [{\citenamefont {Koschorreck}\ \emph {et~al.}(2010)\citenamefont {Koschorreck}, \citenamefont {Napolitano}, \citenamefont {Dubost},\ and\ \citenamefont {Mitchell}}]{PhysRevLett.104.093602}%
  \BibitemOpen
  \bibfield  {author} {\bibinfo {author} {\bibfnamefont {M.}~\bibnamefont {Koschorreck}}, \bibinfo {author} {\bibfnamefont {M.}~\bibnamefont {Napolitano}}, \bibinfo {author} {\bibfnamefont {B.}~\bibnamefont {Dubost}},\ and\ \bibinfo {author} {\bibfnamefont {M.~W.}\ \bibnamefont {Mitchell}},\ }\bibfield  {title} {\bibinfo {title} {Sub-projection-noise sensitivity in broadband atomic magnetometry},\ }\href {https://doi.org/10.1103/PhysRevLett.104.093602} {\bibfield  {journal} {\bibinfo  {journal} {Phys. Rev. Lett.}\ }\textbf {\bibinfo {volume} {104}},\ \bibinfo {pages} {093602} (\bibinfo {year} {2010})}\BibitemShut {NoStop}%
\bibitem [{\citenamefont {Muessel}\ \emph {et~al.}(2014)\citenamefont {Muessel}, \citenamefont {Strobel}, \citenamefont {Linnemann}, \citenamefont {Hume},\ and\ \citenamefont {Oberthaler}}]{PhysRevLett.113.103004}%
  \BibitemOpen
  \bibfield  {author} {\bibinfo {author} {\bibfnamefont {W.}~\bibnamefont {Muessel}}, \bibinfo {author} {\bibfnamefont {H.}~\bibnamefont {Strobel}}, \bibinfo {author} {\bibfnamefont {D.}~\bibnamefont {Linnemann}}, \bibinfo {author} {\bibfnamefont {D.~B.}\ \bibnamefont {Hume}},\ and\ \bibinfo {author} {\bibfnamefont {M.~K.}\ \bibnamefont {Oberthaler}},\ }\bibfield  {title} {\bibinfo {title} {Scalable spin squeezing for quantum-enhanced magnetometry with {Bose-Einstein} condensates},\ }\href {https://doi.org/10.1103/PhysRevLett.113.103004} {\bibfield  {journal} {\bibinfo  {journal} {Phys. Rev. Lett.}\ }\textbf {\bibinfo {volume} {113}},\ \bibinfo {pages} {103004} (\bibinfo {year} {2014})}\BibitemShut {NoStop}%
\bibitem [{\citenamefont {Sewell}\ \emph {et~al.}(2012)\citenamefont {Sewell}, \citenamefont {Koschorreck}, \citenamefont {Napolitano}, \citenamefont {Dubost}, \citenamefont {Behbood},\ and\ \citenamefont {Mitchell}}]{PhysRevLett.109.253605}%
  \BibitemOpen
  \bibfield  {author} {\bibinfo {author} {\bibfnamefont {R.~J.}\ \bibnamefont {Sewell}}, \bibinfo {author} {\bibfnamefont {M.}~\bibnamefont {Koschorreck}}, \bibinfo {author} {\bibfnamefont {M.}~\bibnamefont {Napolitano}}, \bibinfo {author} {\bibfnamefont {B.}~\bibnamefont {Dubost}}, \bibinfo {author} {\bibfnamefont {N.}~\bibnamefont {Behbood}},\ and\ \bibinfo {author} {\bibfnamefont {M.~W.}\ \bibnamefont {Mitchell}},\ }\bibfield  {title} {\bibinfo {title} {Magnetic sensitivity beyond the projection noise limit by spin squeezing},\ }\href {https://doi.org/10.1103/PhysRevLett.109.253605} {\bibfield  {journal} {\bibinfo  {journal} {Phys. Rev. Lett.}\ }\textbf {\bibinfo {volume} {109}},\ \bibinfo {pages} {253605} (\bibinfo {year} {2012})}\BibitemShut {NoStop}%
\bibitem [{\citenamefont {Braunstein}\ and\ \citenamefont {van Loock}(2005)}]{RevModPhys.77.513}%
  \BibitemOpen
  \bibfield  {author} {\bibinfo {author} {\bibfnamefont {S.~L.}\ \bibnamefont {Braunstein}}\ and\ \bibinfo {author} {\bibfnamefont {P.}~\bibnamefont {van Loock}},\ }\bibfield  {title} {\bibinfo {title} {Quantum information with continuous variables},\ }\href {https://doi.org/10.1103/RevModPhys.77.513} {\bibfield  {journal} {\bibinfo  {journal} {Rev. Mod. Phys.}\ }\textbf {\bibinfo {volume} {77}},\ \bibinfo {pages} {513} (\bibinfo {year} {2005})}\BibitemShut {NoStop}%
\bibitem [{sm()}]{sm}%
  \BibitemOpen
  \href@noop {} {\bibinfo  {journal} {See Supplemental Material, which includes Refs.~\cite{BrainPHD,PhysRev.58.1098, PhysRevA.73.062329, PhysRevA.72.052313, PhysRevA.70.044304, hammerer2006quantum}, for the derivations of the theoretical model and technical details of the experiments}\ }\BibitemShut {NoStop}%
\bibitem [{\citenamefont {Leroux}\ \emph {et~al.}(2010)\citenamefont {Leroux}, \citenamefont {Schleier-Smith},\ and\ \citenamefont {Vuleti\ifmmode~\acute{c}\else \'{c}\fi{}}}]{PhysRevLett.104.073602}%
  \BibitemOpen
\bibfield  {journal} {  }\bibfield  {author} {\bibinfo {author} {\bibfnamefont {I.~D.}\ \bibnamefont {Leroux}}, \bibinfo {author} {\bibfnamefont {M.~H.}\ \bibnamefont {Schleier-Smith}},\ and\ \bibinfo {author} {\bibfnamefont {V.}~\bibnamefont {Vuleti\ifmmode~\acute{c}\else \'{c}\fi{}}},\ }\bibfield  {title} {\bibinfo {title} {Implementation of cavity squeezing of a collective atomic spin},\ }\href {https://doi.org/10.1103/PhysRevLett.104.073602} {\bibfield  {journal} {\bibinfo  {journal} {Phys. Rev. Lett.}\ }\textbf {\bibinfo {volume} {104}},\ \bibinfo {pages} {073602} (\bibinfo {year} {2010})}\BibitemShut {NoStop}%
\bibitem [{\citenamefont {Gross}\ \emph {et~al.}(2010)\citenamefont {Gross}, \citenamefont {Zibold}, \citenamefont {Nicklas}, \citenamefont {Estève},\ and\ \citenamefont {Oberthaler}}]{Nature60}%
  \BibitemOpen
  \bibfield  {author} {\bibinfo {author} {\bibfnamefont {C.}~\bibnamefont {Gross}}, \bibinfo {author} {\bibfnamefont {T.}~\bibnamefont {Zibold}}, \bibinfo {author} {\bibfnamefont {E.}~\bibnamefont {Nicklas}}, \bibinfo {author} {\bibfnamefont {J.}~\bibnamefont {Estève}},\ and\ \bibinfo {author} {\bibfnamefont {M.~K.}\ \bibnamefont {Oberthaler}},\ }\bibfield  {title} {\bibinfo {title} {Nonlinear atom interferometer surpasses classical precision limit},\ }\href {https://doi.org/10.1038/nature08919} {\bibfield  {journal} {\bibinfo  {journal} {Nature}\ }\textbf {\bibinfo {volume} {464}},\ \bibinfo {pages} {1165} (\bibinfo {year} {2010})}\BibitemShut {NoStop}%
\bibitem [{\citenamefont {Mao}\ \emph {et~al.}(2023)\citenamefont {Mao}, \citenamefont {Liu}, \citenamefont {Li}, \citenamefont {Cao}, \citenamefont {Chen}, \citenamefont {Xu}, \citenamefont {Tey}, \citenamefont {Huang},\ and\ \citenamefont {You}}]{Nature61}%
  \BibitemOpen
  \bibfield  {author} {\bibinfo {author} {\bibfnamefont {T.-W.}\ \bibnamefont {Mao}}, \bibinfo {author} {\bibfnamefont {Q.}~\bibnamefont {Liu}}, \bibinfo {author} {\bibfnamefont {X.-W.}\ \bibnamefont {Li}}, \bibinfo {author} {\bibfnamefont {J.-H.}\ \bibnamefont {Cao}}, \bibinfo {author} {\bibfnamefont {F.}~\bibnamefont {Chen}}, \bibinfo {author} {\bibfnamefont {W.-X.}\ \bibnamefont {Xu}}, \bibinfo {author} {\bibfnamefont {M.~K.}\ \bibnamefont {Tey}}, \bibinfo {author} {\bibfnamefont {Y.-X.}\ \bibnamefont {Huang}},\ and\ \bibinfo {author} {\bibfnamefont {L.}~\bibnamefont {You}},\ }\bibfield  {title} {\bibinfo {title} {Quantum-enhanced sensing by echoing spin-nematic squeezing in atomic {Bose–Einstein} condensate},\ }\href {https://doi.org/10.1038/s41567-023-02168-3} {\bibfield  {journal} {\bibinfo  {journal} {Nat. Phys.}\ }\textbf {\bibinfo {volume} {19}},\ \bibinfo {pages} {1585} (\bibinfo {year} {2023})}\BibitemShut {NoStop}%
\bibitem [{\citenamefont {Kuzmich}\ \emph {et~al.}(2000)\citenamefont {Kuzmich}, \citenamefont {Mandel},\ and\ \citenamefont {Bigelow}}]{PhysRevLett.85.1594}%
  \BibitemOpen
  \bibfield  {author} {\bibinfo {author} {\bibfnamefont {A.}~\bibnamefont {Kuzmich}}, \bibinfo {author} {\bibfnamefont {L.}~\bibnamefont {Mandel}},\ and\ \bibinfo {author} {\bibfnamefont {N.~P.}\ \bibnamefont {Bigelow}},\ }\bibfield  {title} {\bibinfo {title} {Generation of spin squeezing via continuous quantum nondemolition measurement},\ }\href {https://doi.org/10.1103/PhysRevLett.85.1594} {\bibfield  {journal} {\bibinfo  {journal} {Phys. Rev. Lett.}\ }\textbf {\bibinfo {volume} {85}},\ \bibinfo {pages} {1594} (\bibinfo {year} {2000})}\BibitemShut {NoStop}%
\bibitem [{\citenamefont {Appel}\ \emph {et~al.}(2009)\citenamefont {Appel}, \citenamefont {Windpassinger}, \citenamefont {Oblak}, \citenamefont {Hoff}, \citenamefont {Kj{\ae}rgaard},\ and\ \citenamefont {Polzik}}]{appel2009mesoscopic}%
  \BibitemOpen
  \bibfield  {author} {\bibinfo {author} {\bibfnamefont {J.}~\bibnamefont {Appel}}, \bibinfo {author} {\bibfnamefont {P.~J.}\ \bibnamefont {Windpassinger}}, \bibinfo {author} {\bibfnamefont {D.}~\bibnamefont {Oblak}}, \bibinfo {author} {\bibfnamefont {U.~B.}\ \bibnamefont {Hoff}}, \bibinfo {author} {\bibfnamefont {N.}~\bibnamefont {Kj{\ae}rgaard}},\ and\ \bibinfo {author} {\bibfnamefont {E.~S.}\ \bibnamefont {Polzik}},\ }\bibfield  {title} {\bibinfo {title} {Mesoscopic atomic entanglement for precision measurements beyond the standard quantum limit},\ }\href {https://doi.org/10.1073/pnas.0901550106} {\bibfield  {journal} {\bibinfo  {journal} {Proc. Natl. Acad. Sci.}\ }\textbf {\bibinfo {volume} {106}},\ \bibinfo {pages} {10960} (\bibinfo {year} {2009})}\BibitemShut {NoStop}%
\bibitem [{\citenamefont {Schleier-Smith}\ \emph {et~al.}(2010)\citenamefont {Schleier-Smith}, \citenamefont {Leroux},\ and\ \citenamefont {Vuleti\ifmmode~\acute{c}\else \'{c}\fi{}}}]{PhysRevLett.104.073604}%
  \BibitemOpen
  \bibfield  {author} {\bibinfo {author} {\bibfnamefont {M.~H.}\ \bibnamefont {Schleier-Smith}}, \bibinfo {author} {\bibfnamefont {I.~D.}\ \bibnamefont {Leroux}},\ and\ \bibinfo {author} {\bibfnamefont {V.}~\bibnamefont {Vuleti\ifmmode~\acute{c}\else \'{c}\fi{}}},\ }\bibfield  {title} {\bibinfo {title} {States of an ensemble of two-level atoms with reduced quantum uncertainty},\ }\href {https://doi.org/10.1103/PhysRevLett.104.073604} {\bibfield  {journal} {\bibinfo  {journal} {Phys. Rev. Lett.}\ }\textbf {\bibinfo {volume} {104}},\ \bibinfo {pages} {073604} (\bibinfo {year} {2010})}\BibitemShut {NoStop}%
\bibitem [{\citenamefont {Vasilakis}\ \emph {et~al.}(2015)\citenamefont {Vasilakis}, \citenamefont {Shen}, \citenamefont {Jensen}, \citenamefont {Balabas}, \citenamefont {Salart}, \citenamefont {Chen},\ and\ \citenamefont {Polzik}}]{NP23}%
  \BibitemOpen
  \bibfield  {author} {\bibinfo {author} {\bibfnamefont {G.}~\bibnamefont {Vasilakis}}, \bibinfo {author} {\bibfnamefont {H.}~\bibnamefont {Shen}}, \bibinfo {author} {\bibfnamefont {K.}~\bibnamefont {Jensen}}, \bibinfo {author} {\bibfnamefont {M.}~\bibnamefont {Balabas}}, \bibinfo {author} {\bibfnamefont {D.}~\bibnamefont {Salart}}, \bibinfo {author} {\bibfnamefont {B.}~\bibnamefont {Chen}},\ and\ \bibinfo {author} {\bibfnamefont {E.~S.}\ \bibnamefont {Polzik}},\ }\bibfield  {title} {\bibinfo {title} {Generation of a squeezed state of an oscillator by stroboscopic back-action-evading measurement},\ }\href {https://doi.org/10.1038/nphys3280} {\bibfield  {journal} {\bibinfo  {journal} {Nat. Phys.}\ }\textbf {\bibinfo {volume} {11}},\ \bibinfo {pages} {389} (\bibinfo {year} {2015})}\BibitemShut {NoStop}%
\bibitem [{\citenamefont {Chaudhury}\ \emph {et~al.}(2007)\citenamefont {Chaudhury}, \citenamefont {Merkel}, \citenamefont {Herr}, \citenamefont {Silberfarb}, \citenamefont {Deutsch},\ and\ \citenamefont {Jessen}}]{PhysRevLett.99.163002}%
  \BibitemOpen
  \bibfield  {author} {\bibinfo {author} {\bibfnamefont {S.}~\bibnamefont {Chaudhury}}, \bibinfo {author} {\bibfnamefont {S.}~\bibnamefont {Merkel}}, \bibinfo {author} {\bibfnamefont {T.}~\bibnamefont {Herr}}, \bibinfo {author} {\bibfnamefont {A.}~\bibnamefont {Silberfarb}}, \bibinfo {author} {\bibfnamefont {I.~H.}\ \bibnamefont {Deutsch}},\ and\ \bibinfo {author} {\bibfnamefont {P.~S.}\ \bibnamefont {Jessen}},\ }\bibfield  {title} {\bibinfo {title} {Quantum control of the hyperfine spin of a {Cs} atom ensemble},\ }\href {https://doi.org/10.1103/PhysRevLett.99.163002} {\bibfield  {journal} {\bibinfo  {journal} {Phys. Rev. Lett.}\ }\textbf {\bibinfo {volume} {99}},\ \bibinfo {pages} {163002} (\bibinfo {year} {2007})}\BibitemShut {NoStop}%
\bibitem [{\citenamefont {Fernholz}\ \emph {et~al.}(2008)\citenamefont {Fernholz}, \citenamefont {Krauter}, \citenamefont {Jensen}, \citenamefont {Sherson}, \citenamefont {S\o{}rensen},\ and\ \citenamefont {Polzik}}]{PhysRevLett.101.073601}%
  \BibitemOpen
  \bibfield  {author} {\bibinfo {author} {\bibfnamefont {T.}~\bibnamefont {Fernholz}}, \bibinfo {author} {\bibfnamefont {H.}~\bibnamefont {Krauter}}, \bibinfo {author} {\bibfnamefont {K.}~\bibnamefont {Jensen}}, \bibinfo {author} {\bibfnamefont {J.~F.}\ \bibnamefont {Sherson}}, \bibinfo {author} {\bibfnamefont {A.~S.}\ \bibnamefont {S\o{}rensen}},\ and\ \bibinfo {author} {\bibfnamefont {E.~S.}\ \bibnamefont {Polzik}},\ }\bibfield  {title} {\bibinfo {title} {Spin squeezing of atomic ensembles via nuclear-electronic spin entanglement},\ }\href {https://doi.org/10.1103/PhysRevLett.101.073601} {\bibfield  {journal} {\bibinfo  {journal} {Phys. Rev. Lett.}\ }\textbf {\bibinfo {volume} {101}},\ \bibinfo {pages} {073601} (\bibinfo {year} {2008})}\BibitemShut {NoStop}%
\bibitem [{\citenamefont {Kurucz}\ and\ \citenamefont {M\o{}lmer}(2010)}]{PhysRevA.81.032314}%
  \BibitemOpen
  \bibfield  {author} {\bibinfo {author} {\bibfnamefont {Z.}~\bibnamefont {Kurucz}}\ and\ \bibinfo {author} {\bibfnamefont {K.}~\bibnamefont {M\o{}lmer}},\ }\bibfield  {title} {\bibinfo {title} {Multilevel {Holstein}-{Primakoff} approximation and its application to atomic spin squeezing and ensemble quantum memories},\ }\href {https://doi.org/10.1103/PhysRevA.81.032314} {\bibfield  {journal} {\bibinfo  {journal} {Phys. Rev. A}\ }\textbf {\bibinfo {volume} {81}},\ \bibinfo {pages} {032314} (\bibinfo {year} {2010})}\BibitemShut {NoStop}%
\bibitem [{\citenamefont {Norris}\ \emph {et~al.}(2012)\citenamefont {Norris}, \citenamefont {Trail}, \citenamefont {Jessen},\ and\ \citenamefont {Deutsch}}]{PhysRevLett.109.173603}%
  \BibitemOpen
  \bibfield  {author} {\bibinfo {author} {\bibfnamefont {L.~M.}\ \bibnamefont {Norris}}, \bibinfo {author} {\bibfnamefont {C.~M.}\ \bibnamefont {Trail}}, \bibinfo {author} {\bibfnamefont {P.~S.}\ \bibnamefont {Jessen}},\ and\ \bibinfo {author} {\bibfnamefont {I.~H.}\ \bibnamefont {Deutsch}},\ }\bibfield  {title} {\bibinfo {title} {Enhanced squeezing of a collective spin via control of its qudit subsystems},\ }\href {https://doi.org/10.1103/PhysRevLett.109.173603} {\bibfield  {journal} {\bibinfo  {journal} {Phys. Rev. Lett.}\ }\textbf {\bibinfo {volume} {109}},\ \bibinfo {pages} {173603} (\bibinfo {year} {2012})}\BibitemShut {NoStop}%
\bibitem [{\citenamefont {Kubasik}\ \emph {et~al.}(2009)\citenamefont {Kubasik}, \citenamefont {Koschorreck}, \citenamefont {Napolitano}, \citenamefont {de~Echaniz}, \citenamefont {Crepaz}, \citenamefont {Eschner}, \citenamefont {Polzik},\ and\ \citenamefont {Mitchell}}]{PhysRevA.79.043815}%
  \BibitemOpen
  \bibfield  {author} {\bibinfo {author} {\bibfnamefont {M.}~\bibnamefont {Kubasik}}, \bibinfo {author} {\bibfnamefont {M.}~\bibnamefont {Koschorreck}}, \bibinfo {author} {\bibfnamefont {M.}~\bibnamefont {Napolitano}}, \bibinfo {author} {\bibfnamefont {S.~R.}\ \bibnamefont {de~Echaniz}}, \bibinfo {author} {\bibfnamefont {H.}~\bibnamefont {Crepaz}}, \bibinfo {author} {\bibfnamefont {J.}~\bibnamefont {Eschner}}, \bibinfo {author} {\bibfnamefont {E.~S.}\ \bibnamefont {Polzik}},\ and\ \bibinfo {author} {\bibfnamefont {M.~W.}\ \bibnamefont {Mitchell}},\ }\bibfield  {title} {\bibinfo {title} {Polarization-based light-atom quantum interface with an all-optical trap},\ }\href {https://doi.org/10.1103/PhysRevA.79.043815} {\bibfield  {journal} {\bibinfo  {journal} {Phys. Rev. A}\ }\textbf {\bibinfo {volume} {79}},\ \bibinfo {pages} {043815} (\bibinfo {year} {2009})}\BibitemShut {NoStop}%
\bibitem [{\citenamefont {Hemmer}\ \emph {et~al.}(2021)\citenamefont {Hemmer}, \citenamefont {Monta\~no}, \citenamefont {Baragiola}, \citenamefont {Norris}, \citenamefont {Shojaee}, \citenamefont {Deutsch},\ and\ \citenamefont {Jessen}}]{PhysRevA.104.023710}%
  \BibitemOpen
  \bibfield  {author} {\bibinfo {author} {\bibfnamefont {D.}~\bibnamefont {Hemmer}}, \bibinfo {author} {\bibfnamefont {E.}~\bibnamefont {Monta\~no}}, \bibinfo {author} {\bibfnamefont {B.~Q.}\ \bibnamefont {Baragiola}}, \bibinfo {author} {\bibfnamefont {L.~M.}\ \bibnamefont {Norris}}, \bibinfo {author} {\bibfnamefont {E.}~\bibnamefont {Shojaee}}, \bibinfo {author} {\bibfnamefont {I.~H.}\ \bibnamefont {Deutsch}},\ and\ \bibinfo {author} {\bibfnamefont {P.~S.}\ \bibnamefont {Jessen}},\ }\bibfield  {title} {\bibinfo {title} {Squeezing the angular momentum of an ensemble of complex multilevel atoms},\ }\href {https://doi.org/10.1103/PhysRevA.104.023710} {\bibfield  {journal} {\bibinfo  {journal} {Phys. Rev. A}\ }\textbf {\bibinfo {volume} {104}},\ \bibinfo {pages} {023710} (\bibinfo {year} {2021})}\BibitemShut {NoStop}%
\bibitem [{\citenamefont {Opatrn\'y}\ and\ \citenamefont {M\o{}lmer}(2012)}]{PhysRevA.86.023845}%
  \BibitemOpen
  \bibfield  {author} {\bibinfo {author} {\bibfnamefont {T.}~\bibnamefont {Opatrn\'y}}\ and\ \bibinfo {author} {\bibfnamefont {K.}~\bibnamefont {M\o{}lmer}},\ }\bibfield  {title} {\bibinfo {title} {Spin squeezing and {Schr\"odinger}-cat-state generation in atomic samples with {Rydberg} blockade},\ }\href {https://doi.org/10.1103/PhysRevA.86.023845} {\bibfield  {journal} {\bibinfo  {journal} {Phys. Rev. A}\ }\textbf {\bibinfo {volume} {86}},\ \bibinfo {pages} {023845} (\bibinfo {year} {2012})}\BibitemShut {NoStop}%
\bibitem [{\citenamefont {Gil}\ \emph {et~al.}(2014)\citenamefont {Gil}, \citenamefont {Mukherjee}, \citenamefont {Bridge}, \citenamefont {Jones},\ and\ \citenamefont {Pohl}}]{PhysRevLett.112.103601}%
  \BibitemOpen
  \bibfield  {author} {\bibinfo {author} {\bibfnamefont {L.~I.~R.}\ \bibnamefont {Gil}}, \bibinfo {author} {\bibfnamefont {R.}~\bibnamefont {Mukherjee}}, \bibinfo {author} {\bibfnamefont {E.~M.}\ \bibnamefont {Bridge}}, \bibinfo {author} {\bibfnamefont {M.~P.~A.}\ \bibnamefont {Jones}},\ and\ \bibinfo {author} {\bibfnamefont {T.}~\bibnamefont {Pohl}},\ }\bibfield  {title} {\bibinfo {title} {Spin squeezing in a {Rydberg} lattice clock},\ }\href {https://doi.org/10.1103/PhysRevLett.112.103601} {\bibfield  {journal} {\bibinfo  {journal} {Phys. Rev. Lett.}\ }\textbf {\bibinfo {volume} {112}},\ \bibinfo {pages} {103601} (\bibinfo {year} {2014})}\BibitemShut {NoStop}%
\bibitem [{\citenamefont {Gilmore}\ \emph {et~al.}(2021)\citenamefont {Gilmore}, \citenamefont {Affolter}, \citenamefont {Lewis-Swan}, \citenamefont {Barberena}, \citenamefont {Jordan}, \citenamefont {Rey},\ and\ \citenamefont {Bollinger}}]{gilmore2021quantum}%
  \BibitemOpen
  \bibfield  {author} {\bibinfo {author} {\bibfnamefont {K.~A.}\ \bibnamefont {Gilmore}}, \bibinfo {author} {\bibfnamefont {M.}~\bibnamefont {Affolter}}, \bibinfo {author} {\bibfnamefont {R.~J.}\ \bibnamefont {Lewis-Swan}}, \bibinfo {author} {\bibfnamefont {D.}~\bibnamefont {Barberena}}, \bibinfo {author} {\bibfnamefont {E.}~\bibnamefont {Jordan}}, \bibinfo {author} {\bibfnamefont {A.~M.}\ \bibnamefont {Rey}},\ and\ \bibinfo {author} {\bibfnamefont {J.~J.}\ \bibnamefont {Bollinger}},\ }\bibfield  {title} {\bibinfo {title} {Quantum-enhanced sensing of displacements and electric fields with two-dimensional trapped-ion crystals},\ }\href {https://doi.org/10.1126/science.abi5226} {\bibfield  {journal} {\bibinfo  {journal} {Science}\ }\textbf {\bibinfo {volume} {373}},\ \bibinfo {pages} {673} (\bibinfo {year} {2021})}\BibitemShut {NoStop}%
\bibitem [{\citenamefont {Jin}\ \emph {et~al.}(2024)\citenamefont {Jin}, \citenamefont {Duan}, \citenamefont {Zhang}, \citenamefont {Zhang}, \citenamefont {Bao}, \citenamefont {Shen}, \citenamefont {Xiao}, \citenamefont {Jia}, \citenamefont {Wang},\ and\ \citenamefont {Xiao}}]{PhysRevLett.133.173604}%
  \BibitemOpen
  \bibfield  {author} {\bibinfo {author} {\bibfnamefont {S.}~\bibnamefont {Jin}}, \bibinfo {author} {\bibfnamefont {J.}~\bibnamefont {Duan}}, \bibinfo {author} {\bibfnamefont {Y.}~\bibnamefont {Zhang}}, \bibinfo {author} {\bibfnamefont {X.}~\bibnamefont {Zhang}}, \bibinfo {author} {\bibfnamefont {H.}~\bibnamefont {Bao}}, \bibinfo {author} {\bibfnamefont {H.}~\bibnamefont {Shen}}, \bibinfo {author} {\bibfnamefont {L.}~\bibnamefont {Xiao}}, \bibinfo {author} {\bibfnamefont {S.}~\bibnamefont {Jia}}, \bibinfo {author} {\bibfnamefont {M.}~\bibnamefont {Wang}},\ and\ \bibinfo {author} {\bibfnamefont {Y.}~\bibnamefont {Xiao}},\ }\bibfield  {title} {\bibinfo {title} {Concurrent spin squeezing and light squeezing in an atomic ensemble},\ }\href {https://doi.org/10.1103/PhysRevLett.133.173604} {\bibfield  {journal} {\bibinfo  {journal} {Phys. Rev. Lett.}\ }\textbf {\bibinfo {volume} {133}},\ \bibinfo {pages} {173604} (\bibinfo {year} {2024})}\BibitemShut {NoStop}%
\bibitem [{\citenamefont {Trail}\ \emph {et~al.}(2010)\citenamefont {Trail}, \citenamefont {Jessen},\ and\ \citenamefont {Deutsch}}]{PhysRevLett.105.193602}%
  \BibitemOpen
  \bibfield  {author} {\bibinfo {author} {\bibfnamefont {C.~M.}\ \bibnamefont {Trail}}, \bibinfo {author} {\bibfnamefont {P.~S.}\ \bibnamefont {Jessen}},\ and\ \bibinfo {author} {\bibfnamefont {I.~H.}\ \bibnamefont {Deutsch}},\ }\bibfield  {title} {\bibinfo {title} {Strongly enhanced spin squeezing via quantum control},\ }\href {https://doi.org/10.1103/PhysRevLett.105.193602} {\bibfield  {journal} {\bibinfo  {journal} {Phys. Rev. Lett.}\ }\textbf {\bibinfo {volume} {105}},\ \bibinfo {pages} {193602} (\bibinfo {year} {2010})}\BibitemShut {NoStop}%
\bibitem [{\citenamefont {Bao}\ \emph {et~al.}(2020)\citenamefont {Bao}, \citenamefont {Duan}, \citenamefont {Jin}, \citenamefont {Lu}, \citenamefont {Li}, \citenamefont {Qu}, \citenamefont {Wang}, \citenamefont {Novikova}, \citenamefont {Mikhailov}, \citenamefont {Zhao}, \citenamefont {M{\o}lmer}, \citenamefont {Shen},\ and\ \citenamefont {Xiao}}]{Nature.581.7807}%
  \BibitemOpen
  \bibfield  {author} {\bibinfo {author} {\bibfnamefont {H.}~\bibnamefont {Bao}}, \bibinfo {author} {\bibfnamefont {J.}~\bibnamefont {Duan}}, \bibinfo {author} {\bibfnamefont {S.}~\bibnamefont {Jin}}, \bibinfo {author} {\bibfnamefont {X.}~\bibnamefont {Lu}}, \bibinfo {author} {\bibfnamefont {P.}~\bibnamefont {Li}}, \bibinfo {author} {\bibfnamefont {W.}~\bibnamefont {Qu}}, \bibinfo {author} {\bibfnamefont {M.}~\bibnamefont {Wang}}, \bibinfo {author} {\bibfnamefont {I.}~\bibnamefont {Novikova}}, \bibinfo {author} {\bibfnamefont {E.~E.}\ \bibnamefont {Mikhailov}}, \bibinfo {author} {\bibfnamefont {K.-F.}\ \bibnamefont {Zhao}}, \bibinfo {author} {\bibfnamefont {K.}~\bibnamefont {M{\o}lmer}}, \bibinfo {author} {\bibfnamefont {H.}~\bibnamefont {Shen}},\ and\ \bibinfo {author} {\bibfnamefont {Y.}~\bibnamefont {Xiao}},\ }\bibfield  {title} {\bibinfo {title} {Spin squeezing of $10^{11}$ atoms by prediction and retrodiction measurements},\ }\href {https://doi.org/10.1038/s41586-020-2243-7} {\bibfield  {journal}
  {\bibinfo  {journal} {Nature}\ }\textbf {\bibinfo {volume} {581}},\ \bibinfo {pages} {159} (\bibinfo {year} {2020})}\BibitemShut {NoStop}%
\bibitem [{\citenamefont {Wineland}\ \emph {et~al.}(1994)\citenamefont {Wineland}, \citenamefont {Bollinger}, \citenamefont {Itano},\ and\ \citenamefont {Heinzen}}]{PhysRevA.50.67}%
  \BibitemOpen
  \bibfield  {author} {\bibinfo {author} {\bibfnamefont {D.~J.}\ \bibnamefont {Wineland}}, \bibinfo {author} {\bibfnamefont {J.~J.}\ \bibnamefont {Bollinger}}, \bibinfo {author} {\bibfnamefont {W.~M.}\ \bibnamefont {Itano}},\ and\ \bibinfo {author} {\bibfnamefont {D.~J.}\ \bibnamefont {Heinzen}},\ }\bibfield  {title} {\bibinfo {title} {Squeezed atomic states and projection noise in spectroscopy},\ }\href {https://doi.org/10.1103/PhysRevA.50.67} {\bibfield  {journal} {\bibinfo  {journal} {Phys. Rev. A}\ }\textbf {\bibinfo {volume} {50}},\ \bibinfo {pages} {67} (\bibinfo {year} {1994})}\BibitemShut {NoStop}%
\bibitem [{\citenamefont {Vasilakis}\ \emph {et~al.}(2011)\citenamefont {Vasilakis}, \citenamefont {Shah},\ and\ \citenamefont {Romalis}}]{PhysRevLett.106.143601}%
  \BibitemOpen
  \bibfield  {author} {\bibinfo {author} {\bibfnamefont {G.}~\bibnamefont {Vasilakis}}, \bibinfo {author} {\bibfnamefont {V.}~\bibnamefont {Shah}},\ and\ \bibinfo {author} {\bibfnamefont {M.~V.}\ \bibnamefont {Romalis}},\ }\bibfield  {title} {\bibinfo {title} {Stroboscopic backaction evasion in a dense alkali-metal vapor},\ }\href {https://doi.org/10.1103/PhysRevLett.106.143601} {\bibfield  {journal} {\bibinfo  {journal} {Phys. Rev. Lett.}\ }\textbf {\bibinfo {volume} {106}},\ \bibinfo {pages} {143601} (\bibinfo {year} {2011})}\BibitemShut {NoStop}%
\bibitem [{\citenamefont {Tsang}(2009)}]{PhysRevLett.102.250403}%
  \BibitemOpen
  \bibfield  {author} {\bibinfo {author} {\bibfnamefont {M.}~\bibnamefont {Tsang}},\ }\bibfield  {title} {\bibinfo {title} {Time-symmetric quantum theory of smoothing},\ }\href {https://doi.org/10.1103/PhysRevLett.102.250403} {\bibfield  {journal} {\bibinfo  {journal} {Phys. Rev. Lett.}\ }\textbf {\bibinfo {volume} {102}},\ \bibinfo {pages} {250403} (\bibinfo {year} {2009})}\BibitemShut {NoStop}%
\bibitem [{\citenamefont {Gammelmark}\ \emph {et~al.}(2013)\citenamefont {Gammelmark}, \citenamefont {Julsgaard},\ and\ \citenamefont {M\o{}lmer}}]{PhysRevLett.111.160401}%
  \BibitemOpen
  \bibfield  {author} {\bibinfo {author} {\bibfnamefont {S.}~\bibnamefont {Gammelmark}}, \bibinfo {author} {\bibfnamefont {B.}~\bibnamefont {Julsgaard}},\ and\ \bibinfo {author} {\bibfnamefont {K.}~\bibnamefont {M\o{}lmer}},\ }\bibfield  {title} {\bibinfo {title} {Past quantum states of a monitored system},\ }\href {https://doi.org/10.1103/PhysRevLett.111.160401} {\bibfield  {journal} {\bibinfo  {journal} {Phys. Rev. Lett.}\ }\textbf {\bibinfo {volume} {111}},\ \bibinfo {pages} {160401} (\bibinfo {year} {2013})}\BibitemShut {NoStop}%
\bibitem [{\citenamefont {Zhang}\ and\ \citenamefont {M\o{}lmer}(2017)}]{PhysRevA.96.062131}%
  \BibitemOpen
  \bibfield  {author} {\bibinfo {author} {\bibfnamefont {J.}~\bibnamefont {Zhang}}\ and\ \bibinfo {author} {\bibfnamefont {K.}~\bibnamefont {M\o{}lmer}},\ }\bibfield  {title} {\bibinfo {title} {Prediction and retrodiction with continuously monitored {Gaussian} states},\ }\href {https://doi.org/10.1103/PhysRevA.96.062131} {\bibfield  {journal} {\bibinfo  {journal} {Phys. Rev. A}\ }\textbf {\bibinfo {volume} {96}},\ \bibinfo {pages} {062131} (\bibinfo {year} {2017})}\BibitemShut {NoStop}%
\bibitem [{\citenamefont {Fuderer}\ \emph {et~al.}(2023)\citenamefont {Fuderer}, \citenamefont {Hope},\ and\ \citenamefont {Haine}}]{PhysRevA.108.043722}%
  \BibitemOpen
  \bibfield  {author} {\bibinfo {author} {\bibfnamefont {L.~A.}\ \bibnamefont {Fuderer}}, \bibinfo {author} {\bibfnamefont {J.~J.}\ \bibnamefont {Hope}},\ and\ \bibinfo {author} {\bibfnamefont {S.~A.}\ \bibnamefont {Haine}},\ }\bibfield  {title} {\bibinfo {title} {Hybrid method of generating spin-squeezed states for quantum-enhanced atom interferometry},\ }\href {https://doi.org/10.1103/PhysRevA.108.043722} {\bibfield  {journal} {\bibinfo  {journal} {Phys. Rev. A}\ }\textbf {\bibinfo {volume} {108}},\ \bibinfo {pages} {043722} (\bibinfo {year} {2023})}\BibitemShut {NoStop}%
\bibitem [{\citenamefont {Evrard}\ \emph {et~al.}(2019)\citenamefont {Evrard}, \citenamefont {Makhalov}, \citenamefont {Chalopin}, \citenamefont {Sidorenkov}, \citenamefont {Dalibard}, \citenamefont {Lopes},\ and\ \citenamefont {Nascimbene}}]{PhysRevLett.122.173601}%
  \BibitemOpen
  \bibfield  {author} {\bibinfo {author} {\bibfnamefont {A.}~\bibnamefont {Evrard}}, \bibinfo {author} {\bibfnamefont {V.}~\bibnamefont {Makhalov}}, \bibinfo {author} {\bibfnamefont {T.}~\bibnamefont {Chalopin}}, \bibinfo {author} {\bibfnamefont {L.~A.}\ \bibnamefont {Sidorenkov}}, \bibinfo {author} {\bibfnamefont {J.}~\bibnamefont {Dalibard}}, \bibinfo {author} {\bibfnamefont {R.}~\bibnamefont {Lopes}},\ and\ \bibinfo {author} {\bibfnamefont {S.}~\bibnamefont {Nascimbene}},\ }\bibfield  {title} {\bibinfo {title} {Enhanced magnetic sensitivity with {non-Gaussian} quantum fluctuations},\ }\href {https://doi.org/10.1103/PhysRevLett.122.173601} {\bibfield  {journal} {\bibinfo  {journal} {Phys. Rev. Lett.}\ }\textbf {\bibinfo {volume} {122}},\ \bibinfo {pages} {173601} (\bibinfo {year} {2019})}\BibitemShut {NoStop}%
\bibitem [{\citenamefont {Wang}\ \emph {et~al.}(2017)\citenamefont {Wang}, \citenamefont {Qu}, \citenamefont {Li}, \citenamefont {Bao}, \citenamefont {Vuleti\ifmmode~\acute{c}\else \'{c}\fi{}},\ and\ \citenamefont {Xiao}}]{PhysRevA.96.013823}%
  \BibitemOpen
  \bibfield  {author} {\bibinfo {author} {\bibfnamefont {M.}~\bibnamefont {Wang}}, \bibinfo {author} {\bibfnamefont {W.}~\bibnamefont {Qu}}, \bibinfo {author} {\bibfnamefont {P.}~\bibnamefont {Li}}, \bibinfo {author} {\bibfnamefont {H.}~\bibnamefont {Bao}}, \bibinfo {author} {\bibfnamefont {V.}~\bibnamefont {Vuleti\ifmmode~\acute{c}\else \'{c}\fi{}}},\ and\ \bibinfo {author} {\bibfnamefont {Y.}~\bibnamefont {Xiao}},\ }\bibfield  {title} {\bibinfo {title} {Two-axis-twisting spin squeezing by multipass quantum erasure},\ }\href {https://doi.org/10.1103/PhysRevA.96.013823} {\bibfield  {journal} {\bibinfo  {journal} {Phys. Rev. A}\ }\textbf {\bibinfo {volume} {96}},\ \bibinfo {pages} {013823} (\bibinfo {year} {2017})}\BibitemShut {NoStop}%
\bibitem [{\citenamefont {Luo}\ \emph {et~al.}(2025)\citenamefont {Luo}, \citenamefont {Zhang}, \citenamefont {Chu}, \citenamefont {Maruko}, \citenamefont {Rey},\ and\ \citenamefont {Thompson}}]{luo2024hamiltonianengineeringcollectivexyz}%
  \BibitemOpen
  \bibfield  {author} {\bibinfo {author} {\bibfnamefont {C.}~\bibnamefont {Luo}}, \bibinfo {author} {\bibfnamefont {H.}~\bibnamefont {Zhang}}, \bibinfo {author} {\bibfnamefont {A.}~\bibnamefont {Chu}}, \bibinfo {author} {\bibfnamefont {C.}~\bibnamefont {Maruko}}, \bibinfo {author} {\bibfnamefont {A.~M.}\ \bibnamefont {Rey}},\ and\ \bibinfo {author} {\bibfnamefont {J.~K.}\ \bibnamefont {Thompson}},\ }\bibfield  {title} {\bibinfo {title} {Hamiltonian engineering of collective {XYZ} spin models in an optical cavity},\ }\href {https://doi.org/10.1038/s41567-025-02866-0} {\bibfield  {journal} {\bibinfo  {journal} {Nat. Phys.}\ }\textbf {\bibinfo {volume} {21}},\ \bibinfo {pages} {916} (\bibinfo {year} {2025})}\BibitemShut {NoStop}%
\bibitem [{\citenamefont {Yang}\ \emph {et~al.}(2025)\citenamefont {Yang}, \citenamefont {Luo}, \citenamefont {Zhang}, \citenamefont {Wang}, \citenamefont {Zou}, \citenamefont {Xia},\ and\ \citenamefont {Lu}}]{yang2025minute}%
  \BibitemOpen
  \bibfield  {author} {\bibinfo {author} {\bibfnamefont {Y.}~\bibnamefont {Yang}}, \bibinfo {author} {\bibfnamefont {W.-T.}\ \bibnamefont {Luo}}, \bibinfo {author} {\bibfnamefont {J.-L.}\ \bibnamefont {Zhang}}, \bibinfo {author} {\bibfnamefont {S.-Z.}\ \bibnamefont {Wang}}, \bibinfo {author} {\bibfnamefont {C.-L.}\ \bibnamefont {Zou}}, \bibinfo {author} {\bibfnamefont {T.}~\bibnamefont {Xia}},\ and\ \bibinfo {author} {\bibfnamefont {Z.-T.}\ \bibnamefont {Lu}},\ }\bibfield  {title} {\bibinfo {title} {Minute-scale {Schr{\"o}dinger-cat} state of spin-5/2 atoms},\ }\href {https://doi.org/10.1038/s41566-024-01555-3} {\bibfield  {journal} {\bibinfo  {journal} {Nat. Photonics}\ }\textbf {\bibinfo {volume} {19}},\ \bibinfo {pages} {89} (\bibinfo {year} {2025})}\BibitemShut {NoStop}%
\bibitem [{\citenamefont {Li}\ \emph {et~al.}(2023)\citenamefont {Li}, \citenamefont {Colombo}, \citenamefont {Shu}, \citenamefont {Velez}, \citenamefont {Pilatowsky-Cameo}, \citenamefont {Schmied}, \citenamefont {Choi}, \citenamefont {Lukin}, \citenamefont {Pedrozo-Pe{\~n}afiel},\ and\ \citenamefont {Vuleti{\'c}}}]{li2023improving}%
  \BibitemOpen
  \bibfield  {author} {\bibinfo {author} {\bibfnamefont {Z.}~\bibnamefont {Li}}, \bibinfo {author} {\bibfnamefont {S.}~\bibnamefont {Colombo}}, \bibinfo {author} {\bibfnamefont {C.}~\bibnamefont {Shu}}, \bibinfo {author} {\bibfnamefont {G.}~\bibnamefont {Velez}}, \bibinfo {author} {\bibfnamefont {S.}~\bibnamefont {Pilatowsky-Cameo}}, \bibinfo {author} {\bibfnamefont {R.}~\bibnamefont {Schmied}}, \bibinfo {author} {\bibfnamefont {S.}~\bibnamefont {Choi}}, \bibinfo {author} {\bibfnamefont {M.}~\bibnamefont {Lukin}}, \bibinfo {author} {\bibfnamefont {E.}~\bibnamefont {Pedrozo-Pe{\~n}afiel}},\ and\ \bibinfo {author} {\bibfnamefont {V.}~\bibnamefont {Vuleti{\'c}}},\ }\bibfield  {title} {\bibinfo {title} {Improving metrology with quantum scrambling},\ }\href {https://doi.org/10.1126/science.adg9500} {\bibfield  {journal} {\bibinfo  {journal} {Science}\ }\textbf {\bibinfo {volume} {380}},\ \bibinfo {pages} {1381} (\bibinfo {year} {2023})}\BibitemShut {NoStop}%
\bibitem [{\citenamefont {Colombo}\ \emph {et~al.}(2022)\citenamefont {Colombo}, \citenamefont {Pedrozo-Pe{\~n}afiel}, \citenamefont {Adiyatullin}, \citenamefont {Li}, \citenamefont {Mendez}, \citenamefont {Shu},\ and\ \citenamefont {Vuleti{\'c}}}]{colombo2022time}%
  \BibitemOpen
  \bibfield  {author} {\bibinfo {author} {\bibfnamefont {S.}~\bibnamefont {Colombo}}, \bibinfo {author} {\bibfnamefont {E.}~\bibnamefont {Pedrozo-Pe{\~n}afiel}}, \bibinfo {author} {\bibfnamefont {A.~F.}\ \bibnamefont {Adiyatullin}}, \bibinfo {author} {\bibfnamefont {Z.}~\bibnamefont {Li}}, \bibinfo {author} {\bibfnamefont {E.}~\bibnamefont {Mendez}}, \bibinfo {author} {\bibfnamefont {C.}~\bibnamefont {Shu}},\ and\ \bibinfo {author} {\bibfnamefont {V.}~\bibnamefont {Vuleti{\'c}}},\ }\bibfield  {title} {\bibinfo {title} {Time-reversal-based quantum metrology with many-body entangled states},\ }\href {https://doi.org/10.1038/s41567-022-01653-5} {\bibfield  {journal} {\bibinfo  {journal} {Nat. Phys.}\ }\textbf {\bibinfo {volume} {18}},\ \bibinfo {pages} {925} (\bibinfo {year} {2022})}\BibitemShut {NoStop}%
\bibitem [{\citenamefont {Liu}\ \emph {et~al.}(2022)\citenamefont {Liu}, \citenamefont {Wu}, \citenamefont {Cao}, \citenamefont {Mao}, \citenamefont {Li}, \citenamefont {Guo}, \citenamefont {Tey},\ and\ \citenamefont {You}}]{liu2022nonlinear}%
  \BibitemOpen
  \bibfield  {author} {\bibinfo {author} {\bibfnamefont {Q.}~\bibnamefont {Liu}}, \bibinfo {author} {\bibfnamefont {L.-N.}\ \bibnamefont {Wu}}, \bibinfo {author} {\bibfnamefont {J.-H.}\ \bibnamefont {Cao}}, \bibinfo {author} {\bibfnamefont {T.-W.}\ \bibnamefont {Mao}}, \bibinfo {author} {\bibfnamefont {X.-W.}\ \bibnamefont {Li}}, \bibinfo {author} {\bibfnamefont {S.-F.}\ \bibnamefont {Guo}}, \bibinfo {author} {\bibfnamefont {M.~K.}\ \bibnamefont {Tey}},\ and\ \bibinfo {author} {\bibfnamefont {L.}~\bibnamefont {You}},\ }\bibfield  {title} {\bibinfo {title} {Nonlinear interferometry beyond classical limit enabled by cyclic dynamics},\ }\href {https://doi.org/10.1038/s41567-021-01441-7} {\bibfield  {journal} {\bibinfo  {journal} {Nat. Phys.}\ }\textbf {\bibinfo {volume} {18}},\ \bibinfo {pages} {167} (\bibinfo {year} {2022})}\BibitemShut {NoStop}%
\bibitem [{\citenamefont {Zheng}\ \emph {et~al.}(2022)\citenamefont {Zheng}, \citenamefont {Yang}, \citenamefont {Wang}, \citenamefont {Singh}, \citenamefont {Xiong}, \citenamefont {Xia},\ and\ \citenamefont {Lu}}]{PhysRevLett.129.083001}%
  \BibitemOpen
  \bibfield  {author} {\bibinfo {author} {\bibfnamefont {T.~A.}\ \bibnamefont {Zheng}}, \bibinfo {author} {\bibfnamefont {Y.~A.}\ \bibnamefont {Yang}}, \bibinfo {author} {\bibfnamefont {S.-Z.}\ \bibnamefont {Wang}}, \bibinfo {author} {\bibfnamefont {J.~T.}\ \bibnamefont {Singh}}, \bibinfo {author} {\bibfnamefont {Z.-X.}\ \bibnamefont {Xiong}}, \bibinfo {author} {\bibfnamefont {T.}~\bibnamefont {Xia}},\ and\ \bibinfo {author} {\bibfnamefont {Z.-T.}\ \bibnamefont {Lu}},\ }\bibfield  {title} {\bibinfo {title} {Measurement of the electric dipole moment of $^{171}\mathrm{Yb}$ atoms in an optical dipole trap},\ }\href {https://doi.org/10.1103/PhysRevLett.129.083001} {\bibfield  {journal} {\bibinfo  {journal} {Phys. Rev. Lett.}\ }\textbf {\bibinfo {volume} {129}},\ \bibinfo {pages} {083001} (\bibinfo {year} {2022})}\BibitemShut {NoStop}%
\bibitem [{\citenamefont {Lysne}\ \emph {et~al.}(2020)\citenamefont {Lysne}, \citenamefont {Kuper}, \citenamefont {Poggi}, \citenamefont {Deutsch},\ and\ \citenamefont {Jessen}}]{PhysRevLett.124.230501}%
  \BibitemOpen
  \bibfield  {author} {\bibinfo {author} {\bibfnamefont {N.~K.}\ \bibnamefont {Lysne}}, \bibinfo {author} {\bibfnamefont {K.~W.}\ \bibnamefont {Kuper}}, \bibinfo {author} {\bibfnamefont {P.~M.}\ \bibnamefont {Poggi}}, \bibinfo {author} {\bibfnamefont {I.~H.}\ \bibnamefont {Deutsch}},\ and\ \bibinfo {author} {\bibfnamefont {P.~S.}\ \bibnamefont {Jessen}},\ }\bibfield  {title} {\bibinfo {title} {Small, highly accurate quantum processor for intermediate-depth quantum simulations},\ }\href {https://doi.org/10.1103/PhysRevLett.124.230501} {\bibfield  {journal} {\bibinfo  {journal} {Phys. Rev. Lett.}\ }\textbf {\bibinfo {volume} {124}},\ \bibinfo {pages} {230501} (\bibinfo {year} {2020})}\BibitemShut {NoStop}%
\bibitem [{\citenamefont {Liu}\ \emph {et~al.}(2025)\citenamefont {Liu}, \citenamefont {Xue}, \citenamefont {Radzihovsky}, \citenamefont {Li}, \citenamefont {Vasilyev}, \citenamefont {Wu},\ and\ \citenamefont {Vuleti\ifmmode~\acute{c}\else \'{c}\fi{}}}]{25ds-9724}%
  \BibitemOpen
  \bibfield  {author} {\bibinfo {author} {\bibfnamefont {Q.}~\bibnamefont {Liu}}, \bibinfo {author} {\bibfnamefont {M.}~\bibnamefont {Xue}}, \bibinfo {author} {\bibfnamefont {M.}~\bibnamefont {Radzihovsky}}, \bibinfo {author} {\bibfnamefont {X.}~\bibnamefont {Li}}, \bibinfo {author} {\bibfnamefont {D.~V.}\ \bibnamefont {Vasilyev}}, \bibinfo {author} {\bibfnamefont {L.-N.}\ \bibnamefont {Wu}},\ and\ \bibinfo {author} {\bibfnamefont {V.}~\bibnamefont {Vuleti\ifmmode~\acute{c}\else \'{c}\fi{}}},\ }\bibfield  {title} {\bibinfo {title} {Enhancing dynamic range of sub-standard-quantum-limit measurements via quantum deamplification},\ }\href {https://doi.org/10.1103/25ds-9724} {\bibfield  {journal} {\bibinfo  {journal} {Phys. Rev. Lett.}\ }\textbf {\bibinfo {volume} {135}},\ \bibinfo {pages} {040801} (\bibinfo {year} {2025})}\BibitemShut {NoStop}%
\bibitem [{\citenamefont {Julsgaard}(2003)}]{BrainPHD}%
  \BibitemOpen
  \bibfield  {author} {\bibinfo {author} {\bibfnamefont {B.}~\bibnamefont {Julsgaard}},\ }\href@noop {} {\bibfield  {journal} {\bibinfo  {journal} {Entanglement and quantum interactions with macroscopic gas samples, Ph.D. thesis, University of Aarhus}\ } (\bibinfo {year} {2003})}\BibitemShut {NoStop}%
\bibitem [{\citenamefont {Holstein}\ and\ \citenamefont {Primakoff}(1940)}]{PhysRev.58.1098}%
  \BibitemOpen
  \bibfield  {author} {\bibinfo {author} {\bibfnamefont {T.}~\bibnamefont {Holstein}}\ and\ \bibinfo {author} {\bibfnamefont {H.}~\bibnamefont {Primakoff}},\ }\bibfield  {title} {\bibinfo {title} {Field dependence of the intrinsic domain magnetization of a ferromagnet},\ }\href {https://doi.org/10.1103/PhysRev.58.1098} {\bibfield  {journal} {\bibinfo  {journal} {Phys. Rev.}\ }\textbf {\bibinfo {volume} {58}},\ \bibinfo {pages} {1098} (\bibinfo {year} {1940})}\BibitemShut {NoStop}%
\bibitem [{\citenamefont {Muschik}\ \emph {et~al.}(2006)\citenamefont {Muschik}, \citenamefont {Hammerer}, \citenamefont {Polzik},\ and\ \citenamefont {Cirac}}]{PhysRevA.73.062329}%
  \BibitemOpen
  \bibfield  {author} {\bibinfo {author} {\bibfnamefont {C.~A.}\ \bibnamefont {Muschik}}, \bibinfo {author} {\bibfnamefont {K.}~\bibnamefont {Hammerer}}, \bibinfo {author} {\bibfnamefont {E.~S.}\ \bibnamefont {Polzik}},\ and\ \bibinfo {author} {\bibfnamefont {J.~I.}\ \bibnamefont {Cirac}},\ }\bibfield  {title} {\bibinfo {title} {Efficient quantum memory and entanglement between light and an atomic ensemble using magnetic fields},\ }\href {https://doi.org/10.1103/PhysRevA.73.062329} {\bibfield  {journal} {\bibinfo  {journal} {Phys. Rev. A}\ }\textbf {\bibinfo {volume} {73}},\ \bibinfo {pages} {062329} (\bibinfo {year} {2006})}\BibitemShut {NoStop}%
\bibitem [{\citenamefont {Hammerer}\ \emph {et~al.}(2005)\citenamefont {Hammerer}, \citenamefont {Polzik},\ and\ \citenamefont {Cirac}}]{PhysRevA.72.052313}%
  \BibitemOpen
  \bibfield  {author} {\bibinfo {author} {\bibfnamefont {K.}~\bibnamefont {Hammerer}}, \bibinfo {author} {\bibfnamefont {E.~S.}\ \bibnamefont {Polzik}},\ and\ \bibinfo {author} {\bibfnamefont {J.~I.}\ \bibnamefont {Cirac}},\ }\bibfield  {title} {\bibinfo {title} {Teleportation and spin squeezing utilizing multimode entanglement of light with atoms},\ }\href {https://doi.org/10.1103/PhysRevA.72.052313} {\bibfield  {journal} {\bibinfo  {journal} {Phys. Rev. A}\ }\textbf {\bibinfo {volume} {72}},\ \bibinfo {pages} {052313} (\bibinfo {year} {2005})}\BibitemShut {NoStop}%
\bibitem [{\citenamefont {Hammerer}\ \emph {et~al.}(2004)\citenamefont {Hammerer}, \citenamefont {M\o{}lmer}, \citenamefont {Polzik},\ and\ \citenamefont {Cirac}}]{PhysRevA.70.044304}%
  \BibitemOpen
  \bibfield  {author} {\bibinfo {author} {\bibfnamefont {K.}~\bibnamefont {Hammerer}}, \bibinfo {author} {\bibfnamefont {K.}~\bibnamefont {M\o{}lmer}}, \bibinfo {author} {\bibfnamefont {E.~S.}\ \bibnamefont {Polzik}},\ and\ \bibinfo {author} {\bibfnamefont {J.~I.}\ \bibnamefont {Cirac}},\ }\bibfield  {title} {\bibinfo {title} {Light-matter quantum interface},\ }\href {https://doi.org/10.1103/PhysRevA.70.044304} {\bibfield  {journal} {\bibinfo  {journal} {Phys. Rev. A}\ }\textbf {\bibinfo {volume} {70}},\ \bibinfo {pages} {044304} (\bibinfo {year} {2004})}\BibitemShut {NoStop}%
\bibitem [{\citenamefont {Hammerer}(2006)}]{hammerer2006quantum}%
  \BibitemOpen
  \bibfield  {author} {\bibinfo {author} {\bibfnamefont {K.}~\bibnamefont {Hammerer}},\ }\href@noop {} {\bibfield  {journal} {\bibinfo  {journal} {Quantum information processing with atomic ensembles and light, Ph.D. thesis, Technische Universit{\"a}t M{\"u}nchen}\ } (\bibinfo {year} {2006})}\BibitemShut {NoStop}%
\end{thebibliography}%
\end{document}